\documentclass[pra,twocolumn,showpacs,superscriptaddress,amsmath]{revtex4}

\usepackage[utf8]{inputenc}
\usepackage[german,american]{babel}
\usepackage{times}
\usepackage[T1]{fontenc}
\usepackage{bm}
\usepackage{amssymb}
\usepackage{graphicx}
\usepackage{epsfig}
\usepackage{color}
\usepackage{ulem}

\newcommand{\hide}[1]{}

\newcommand{\eq}[1]{Eq.\,(\ref{#1})}
\newcommand{\eqs}[1]{Eqs.\,(\ref{#1})}
\newcommand{\noeq}[1]{(\ref{#1})}
\newcommand{\fig}[1]{Fig.\,\ref{#1}}

\newcommand{\bra}[1]{\langle \rm #1 | \,}
\newcommand{\ket}[1]{\, | \rm #1 \rangle}

\renewcommand{\a}{\hat a}
\newcommand{\ad}{\hat a^\dagger}

\newcommand{\mS}{\mathcal{S}}
\newcommand{\mtS}{\tilde{\mathcal{S}}}

\newcommand{\na}{\hat{n}}

\usepackage{hyperref}

\renewcommand{\BibitemShut}[1]{}

\begin{document}

\title{Interacting in-plane molecular dipoles in a zig-zag chain}

\author{Qingyang Wang}
\thanks{These authors contributed equally.}
\affiliation{Department of Physics, Massachusetts Institute of Technology, Cambridge, MA  02139, USA}
\affiliation{Department of Physics, Harvard University, Cambridge, MA 02138, USA}

\author{Johannes Otterbach$^*$}
\affiliation{Department of Physics, Harvard University, Cambridge, MA 02138, USA}

\author{Susanne F. Yelin}
\affiliation{Department of Physics, Harvard University, Cambridge, MA 02138, USA}
\affiliation{Department of Physics, University of Connecticut, Storrs, CT 06269, USA}

\date{\today}

\begin{abstract}
The system with externally polarized dipole molecules at half-filling moving along a one-dimensional zig-zag chain is studied, including the ground-state phase diagram. The dipoles are oriented in-plane. Together with the geometry of the chain this gives rise to a bond-alternating nearest neighbor interaction due to simultaneous attractive and repulsive interactions. Because of the quantum Zeno effect due to the reactive nature of molecules \hide{, along with energy and entropy arguments,} the system can be treated as hard-core. By tuning the ratio between the nearest-neighbor interaction and hopping, various phases can be accessed by controlling the polarization angle. In the ultra-strong coupling limit, the system simplifies to a frustrated extended axial Ising model. \hide{An exact phase diagram is shown in this limit.} For the  small coupling limit, qualitative discussion of the ordering behavior using effective field theory arguments is provided. We show that when chain angle is small, the system mostly exhibits BKT-type phase transitions, whereas large chain angle would drive the system into a gapped (Ising) dimerized phase, where the hopping strength is closely related to the orientation of dimerized pairs. 
\end{abstract}

\pacs{}

\maketitle

\section{INTRODUCTION}
The efficient production of ultra-cold dipolar systems has paved the way to a wide range of interesting effects, for example, strongly correlated systems, chemical reactions at ultracold temperatures, precision tests of fundamental symmetries, possibly new scheme of quantum information processing, just to mention a few \cite{Carr2009, Baranov2012}.  Additionally, there has been great progress in the creation of new techniques for non-standard optical lattices \cite{Becker2010,Windpassinger2013} and optical tweezers \cite{Nogrette2014a} that would make a quantum simulator using ultracold atoms systems even more promising and unique. The vast tunability offered by molecules and lattice configuration has introduced many ideas to simulate interesting unsolved quantum models motivated by solid-state physics. In particular, low dimensional systems in this context are of great interest, partly because of the recent development in creating real solid state systems that can be described in theoretical models studied in the past, and also because an ultracold system may provide a test ground that is beyond the actual material we have access to today. Topics in low dimensional physics range from frustrated systems in 1D, 2D \cite{Ohgoe2012,Albuquerque2011, Mezzacapo2012,Varney2011,Micheli2006,Gorshkov2013}, and coupled one-dimensional setups \cite{Sherkunov2012,Knap2012,Volosniev2013, Bauer2012}, to non-equilibrium behavior in certain systems \cite{Trotzky2011c,Hofferberth2007a}. 

With this, we consider a quasi-1D system, where the dipolar particles are confined in a zig-zag  optical lattice and are polarized in-plane, leading to simultaneous attractive and repulsive interactions (\fig{fig:model}). Depending on the angle of the zig-zag opening, this model can be viewed as the 1D building block of, for instance, a hexagonal or kagom\'e lattice. We qualitatively study the  phase diagram (\fig{fig:phasediagram}), in particular for the two limiting cases -- very small and very large inter-site hopping. Fundamentally, this model can incorporate essential ingredients of 1D frustrated systems, namely long range coupling and/or long-range hopping with great flexibility, therefore making it a promising model for the study of quasi-1D frustrated systems in particular.  This model can, in principle, be explored with most of species of polar particles.

It should be pointed out that complementary studies have been done on the locally interacting Hubbard chain with next-nearest neighbor hopping \cite{Japaridze2007} and the frustrated triangular lattice with nearest-neighbor interactions \cite{Mishra2013}. The dynamics of dipoles confined to two independent chains with weak intra-chain hopping was analyzed in the paper \cite{Gammelmark2013}.


\section{THE MODEL}
\begin{figure}
\includegraphics[scale=0.35]{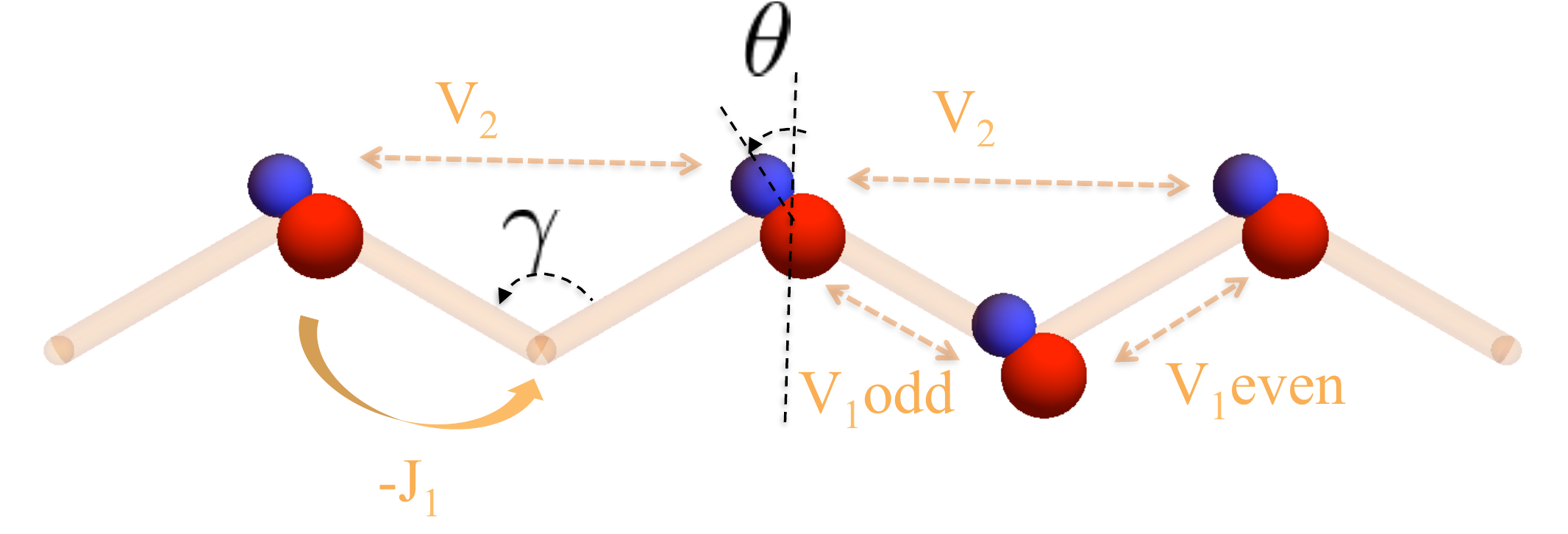}
\caption{\label{fig:model} (Color online) Schematic setup of  dipoles moving on a zig-zag chain with an opening angle $\gamma$. The dipoles are polarized by an external field enclosing an angle $\theta$ with the normal of the chain axis }
\end{figure}
Throughout the paper, we set the temperature to be zero. The model we consider is conceptually described in \fig{fig:model}. This system consists of hard-core dipoles sitting at the vertices of the zig-zag chain with chain opening angle $\gamma$ ($0<\gamma \leq \pi$, cf. \fig{fig:model}). The dipoles can be realized using heteronuclear molecules \cite{Ni2008,Takekoshi2012,Yan2013} or dipolar atoms \cite{Lahaye2007,Chotia2012,Gorshkov2013,Baranov2012}. The dipolar particles are polarized in-plane, leading to simultaneous attractive and repulsive interactions from dipole-dipole interactions 
\[
V_{dip}=\epsilon_{dd}(1-3\cos^2 \theta_{\bf{r_1}-\bf{r_2}})
\]
 with the dipolar coupling strength $\epsilon_{dd}=\mu_e /(4\pi \epsilon_0|\bf{r_1}-\bf{r_2}|^3)$, where $\epsilon_0$, $\mu_e $ are the vacuum permittivity and electric dipole moment of the molecules, respectively, $ \bf{r_1}$ and $\bf{r_2}$ are the position of the molecules, $\theta_{\bf{r_1}-\bf{r_2}}$ is the angle between ($\bf{r_1}-\bf{r_2}$) and the external electric field that polarizes the molecules. Additionally the particles are mobile and can propagate along a zig-zag chain. The system thus represents the edge of a honeycomb lattice or a two-leg ladder with suppressed intra-chain hopping.
 $\epsilon_{dd}=\mu_e /4\pi \epsilon_0|\bf{r_1}-\bf{r_2}|^3$

In actual experiments, the lattice can be created by appropriately angled standing wave laser fields with the correct intensities to create single-chain strands. The molecules can then be loaded by applying an electric field $\vec E$ perpendicular to the zig-zag plane, and subsequently changing the orientation of $\vec E$ adiabatically until it becomes parallel to the plane, followed by the process of changing $\vec E$ in plane (to vary $\theta$). This way, there should never be more than one molecule per site, fulfilling the hardcore condition throughout the experiment (see next subsection for more detail). 

The most general Hamiltonian that describes our system is
\begin{align}
 H =&  -\sum_{j'>j}\sum_{j}J_{j'-j} \ad_j\a_{j'} +\text{h.c.}\nonumber \\
 &-\mu\sum_{j} \na_j  + \sum_{j'>j} \sum_{j} V^{[j/2]}_{j'-j} \ \na_j\na_{j'}
\label{eq:general}
\end{align}
where $J_{j-j'}$ is the hopping parameter between sites $j$ and $j'$, $\mu$ is the chemical potential. Note that the creation (destruction) operators $\a_{j} (\ad_{j}$) can either be fermionic or bosonic without having any essential difference as there is a exact mapping from fermion to hardcore boson systems \cite{Lieb1963} in this case. $V^{[j/2]}_{j-j'}$ denotes the non-local dipole-dipole interactions between particles at site $j$ and $j'$, respectively. Note that due to the anisotropic nature of dipole-dipole interaction and the non-trivial geometry of the chain, this interaction term $V^{[j/2]}_{j-j'}$ depends not only on the range $j-j'$ but also on the even-odd of $j$ (expressed by $[j/2]$). This $V^{[j/2]}_{j-j'}$ can be varied dynamically from negative to positive value with $\theta$ and $\gamma$. As an example, using the standard form of the dipole interaction, we find, after simple trigonometric manipulations, the following explicit expressions for the nearest neighbor (NN) interaction and next nearest neighbor (NNN) interaction.
\begin{align}
 V_1^{\text{even}} \,=\,& \epsilon_{dd} \left[1-3\cos^2\left(\pi-\frac{\gamma}{2}-\theta\right)\right], \\
 V_1^{\text{odd}} \,=\,& \epsilon_{dd} \left[1-3\cos^2\left(\frac{\gamma}{2}-\theta\right)\right], \\
 V_{2} \,=\,&  \frac{\epsilon_{dd}}{\left[2(1-\cos(\gamma)) \right]^{3/2}} \left[1-3\cos^2\left(\frac{\pi}{2}-\theta\right)\right].
\end{align}

\subsection*{Simplification of the Hamiltonian}
We simplify the model \eq{eq:general} by assuming that there are exactly half as many molecules as lattice sites. This is a somewhat less specific assumption than it looks at first glance, since the remaining parameters can be mostly rescaled for relatively small filling imbalances. In addition, we further impose that the lattice opening angle $\gamma \geq 2\pi/3 $. This allows us to safely ignore longer-range hopping (beyond $J_1$, i.e., NN hopping) as the overlap between the next-nearest-neighbor Wannier orbitals and beyond is significantly smaller than the nearest-neighbor ones. Likewise, we make the simplification on the (dipolar) interaction terms by only taking NN and NNN interactions. All the contribution from longer-range interaction is small because of the $1/r^3$ nature of dipolar interaction and may be ignored \footnote{There are questions such as many-body localization where there can be a difference between this simple model of ``mid-range'' interactions and true long-range interaction. But none of these are touched upon in this project.}. 
With this we introduce the dimerization parameter $\delta$ and thus the Hamiltonian of the system is reduced to
\begin{eqnarray}
 H &=& -J_1\sum_{j} \ \ad_{j+1}\a_{j}+\text{h.c.}\nonumber \\ 
 &&+ V_{\rm NN}\sum_{j} [1+\delta(-1)^j] \na_j\na_{j+1}  \nonumber + V_2 \sum_{j}  \ \na_j\na_{j+2} \\
  &&-\mu\sum_{j} \na_j 
 \label{eq:main}
\end{eqnarray}  

where $V_{\rm NN}$ and $\delta$ are related to $V_1^{\text{even}}$ and $V_1^{\text{odd}}$ as $\delta=(V_1^{\text{even}}-V_1^{\text{odd}})/(V_1^{\text{odd}}+V_1^{\text{even}})$ and $V_{\rm NN}=(V_1^{\text{odd}}+V_1^{\text{even}})/2$. In this paper we study the model described by this Hamiltonian. 

\begin{figure}[h]
\includegraphics[scale=0.55]{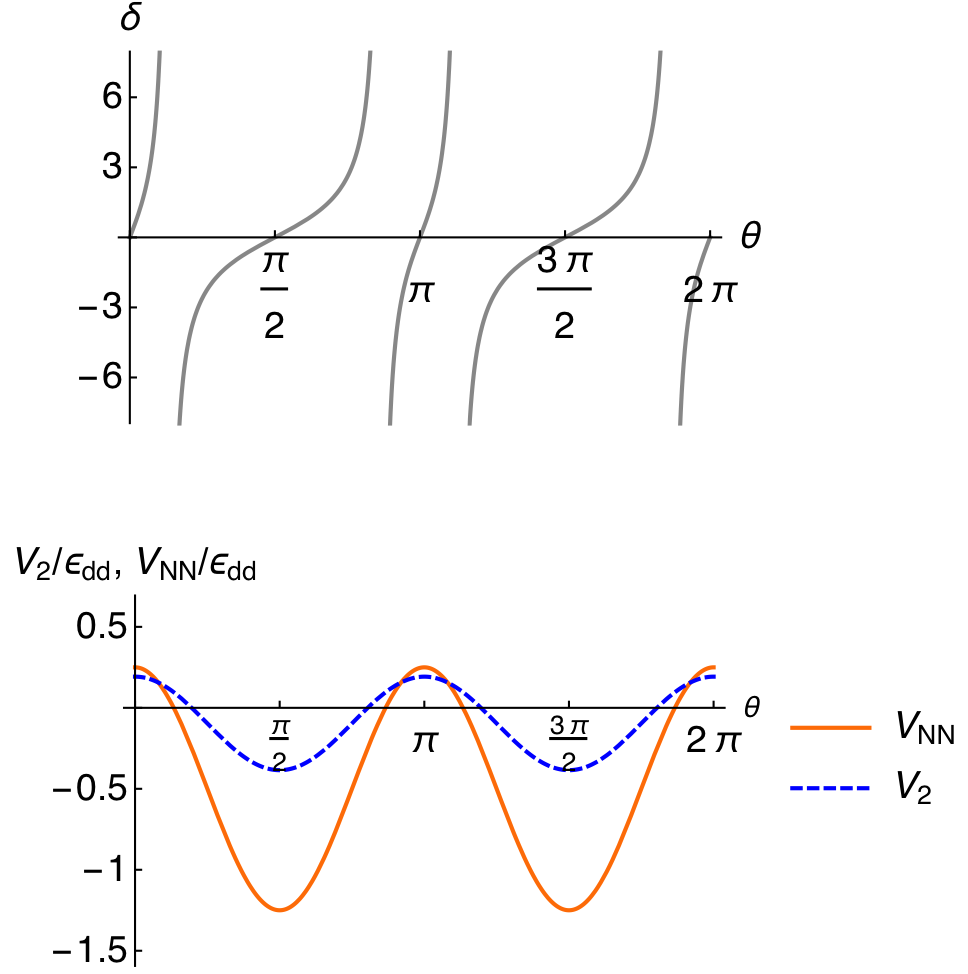}
\caption{\label{fig:dimpara} (Color online) Dimerization parameter $\delta$ and interactions $V_{\rm NN}$, $V_2$ with respect to $\theta$. The chain opening angle $\gamma$ is set to $\gamma = 2\pi/3$.}
\end{figure}

We restrict the region of parameters $\theta$ and $\delta$ by symmetry arguments. First, we note that the interactions exhibit symmetries with respect to $\theta = 0$ (cf. \fig{fig:dimpara}), which translate directly into symmetries of the Hamiltonian. Performing the transformation $\theta\rightarrow \theta+\pi$ leaves the Hamiltonian Eq.(\ref{eq:main}) unchanged and we can restrict ourselves to the range $\theta \in [-\pi/2,\pi/2]$. Another symmetry is changing the sign of the dimerization parameter as $\delta\rightarrow - \delta$ while at the same time inverting theta $\theta\rightarrow -\theta$. However inverting the sign of $\delta$ can be achieved merely by shifting the summation index by $\pm 1$. Therefore we can further restrict ourself to $\delta>0$ and $\theta \in [0,\pi/2]$. This implies that the dull translational invariance is broken, yet as we will see shortly, these symmetries will be reproduced in the systems' ground-states.

\subsection*{On-site contribution and stability}

 In general, the models of polar molecules in optical lattices come with the on-site interaction term  $Un_i(n_i-1)/2$. This term is often abandoned when the molecules are polarized by an external electric field and thus can be regarded as hard core bosons. This is because of the large on-site energy generated by two dipoles feeling strong repulsive force being close at the same site, essentially treating $U \to \infty$. This simplification process, however, needs extra care in our case since, once the E field id in plane, the orientation of dipoles changes between strong attractive and strong repulsive interactions, depending on the (in-plane) polarization angle.  Here we argue that in most of cases the on-site term can still be ignored mainly because of the quantum Zeno effect.

 To go into a little more detail, we first give an estimate on the on-site interaction energy $U$. This is computed as 
\begin{align}
U &= U_{ct}+U_{dip} \\ \nonumber
&=g\int d^3r \rho(\bm{r})^2 + \int d^3r d^3r' \rho(\bm{r})U_{dd}(\bm{r}-\bm{r}')\rho(\bm{r}')
\end{align}
where the first term is the effective contact potential and the second is the potential coming from the dipole interaction. $\rho(\bf{r})=|w(\bf{r})|^2$ is the Wannier function density, $U_{dd}$ is the dipolar interaction, and $g$ is the depth of contact potential that is related to s-wave scattering length. The second term is expressed in Fourier-transformed $\tilde{\rho}$ and $\tilde{U}_{dd}$ as, $1/(2\pi)^2\int d^2 \bm{k} \tilde{\rho}(\bm{k})^2 \tilde{U_{dd}}(\bm{k})$  Here we assume a strong trapping potential in the $z$-direction, thereby treating the lattice site as 2D, and further assume that for each site the trap is isotropic. The polarizing E field is also in this plane, and \hide{since the trap is isotropic} thus the direction of E field in the $xy$-plane is irrelevant in the discussion. If treating Wannier functions as Gaussians with length scale $l_{\rm HO}$, then $\tilde{\rho}(\bm{k})=\exp{(-l_{\rm HO}^2 k^2/4)}$, and $\tilde{U_{dd}}(\bm{k})= -\pi d^2(1/\epsilon-k)+\pi d^2 q \cos{(2\phi_k)}$, where $d$ is the electric dipole moment and $\epsilon$ is the cutoff length that is on the order of molecule length. From this we arrive  at
\begin{align}
U_{dip} &=\int d\bm{k}^2\left[-\pi d^2\left(\frac{1}{\epsilon}-q\right)+\pi d^2 q \cos{(2\phi_k)} \right]e^{-\frac{1}{2}l_{\rm HO}^2k^2} \nonumber \\
&=\frac{2\pi^2 d^2}{l_{\rm HO}^2}\left(\frac{\sqrt{2\pi}}{l_{\rm HO}} -\frac{1}{\epsilon}\right)
\end{align}
Typically, $l_{\rm HO}\approx 1000 \text{nm}$ and $\epsilon \approx 0.1 \text{nm}$. Therefore the on-site energy $U$ is negative with an absolute value at least several orders of magnitude larger than the other energy scales such as $J_1$, $V_{\rm NN}$ and $V_2$, which are at most on the order of $d^2/l_{\rm HO}^3$. If we naively ignore the dynamics and internal structure of the molecules and assume the system initially is prepared with one molecule per site at most, we can neglect the part of the Hilbert space with more than one molecule per site. This can be done because in the ultracold regime, there would be no process to dissipate the energy gained from this on-site contribution.


Often however, the molecules are reactive and hence will be kicked out of the optical lattice once they come to occupy the same lattice site. In these situations, attractive dipole directions enhance such reactive processes and and the appropriate dissipative picture is necessary to describe those systems. This is in contrast to the case where molecules are polarized to be repulsive and consequently feel the huge potential barrier generated by the dipole interactions before they can approach close enough to start inelastic processes. Even with the dissipation process, we point out that when the dissipation is strong, the decay process of molecules is frozen out. This counter-intuitive result is due to the continuous Zeno effect \cite{Paredes2004a, Yan2013, GarciaRipoll2009a}. When $\gamma \gg J$, where $\gamma$ is the 2-body on-site loss rate and $J$ is the hopping parameter, the molecules may again be treated as hard-core, with much slower dissipation rate of the system $\gamma_{eff} \approx J^2/\gamma \ll 1/J$. Thus, it is necessary to choose the system parameters such that the tricky cases are avoided. In what follows this is assumed.

\section{ULTRASTRONG COUPLING CASE}
In this section we consider first the ultra-strong coupling limit $J_1 \to 0$ with an even number of particles, i.e. $N\in 2\mathbb{N}$, where we observe that Hamiltonian (\ref{eq:main}) reduces to a purely classical one. We project the system onto a spin-$1/2$ system where the spin degree of freedom is encoded in the occupation number of a single lattice site, which is explicitly done by the Jordan-Wigner transformation, $S^+_j=a^\dag_j e^{i\pi O_j}, \, S^-_j=a_j e^{-i\pi O_j}, \, S^z_j=a_j^\dag a_j - \frac{1}{2}$, with
$O_j=\sum_{l<j}a_l^\dag a_l$. Here the $S^+$ and $S^-$ operators are spin raising and lowering operators respectively.

\begin{figure*}
\includegraphics[scale=0.40]{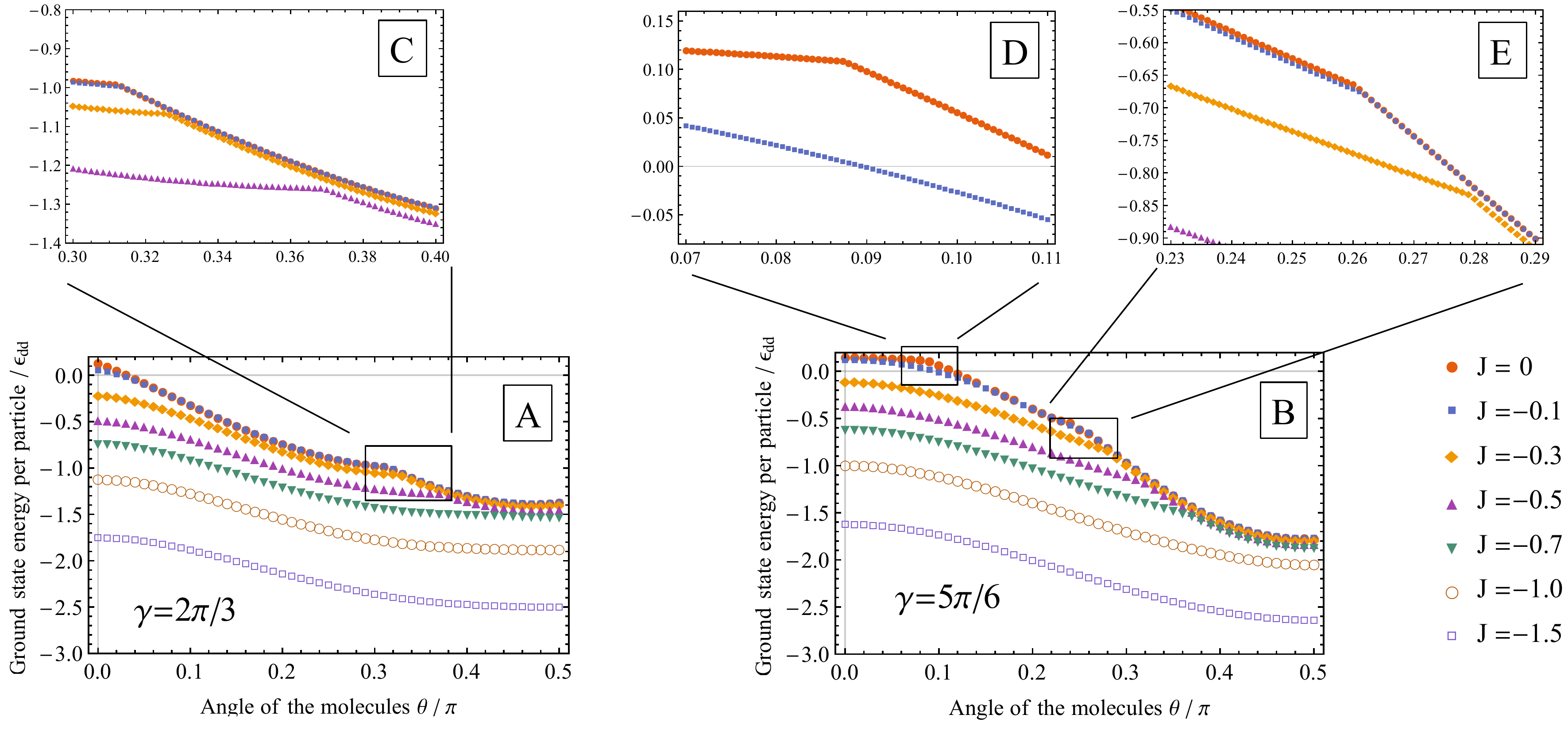}
\caption{\label{fig:sum2} (Color online) Ground state energy per particle plotted against $\theta \in [-\pi/2,\pi/2]$ with various hopping parameter $J_1$ by exact diagnolization method. Number of sites L=18. Left: $\gamma=2\pi/3$ \, Right: $\gamma=5\pi/6$. The kinks show the first order phase transition points.}
\end{figure*}

\begin{figure}[h]
\includegraphics[scale=0.60]{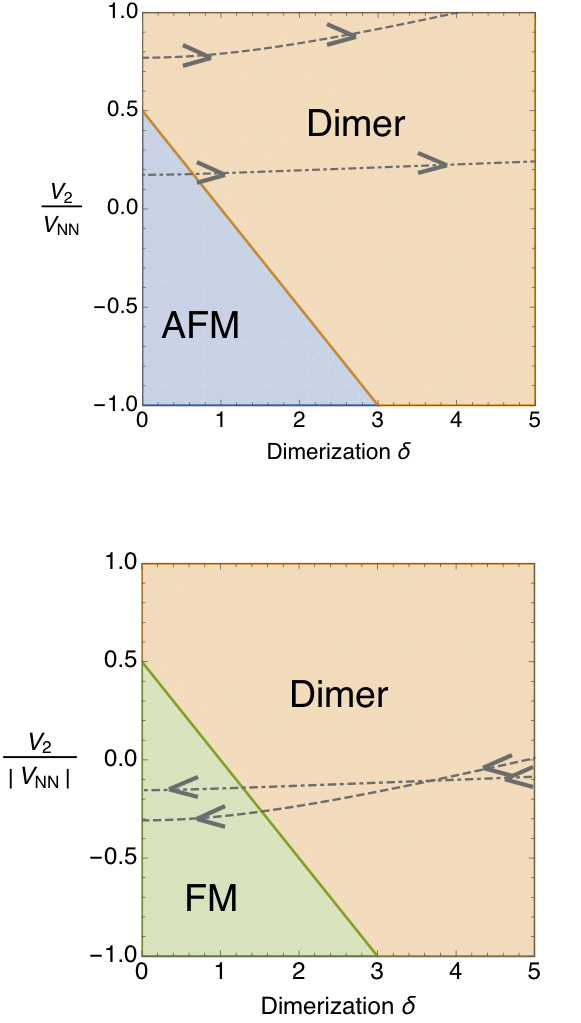}[h]
\caption{ (Color online) Phase diagram of the ultrastrong coupling limit. Top: $V_{\rm NN}>0$ \, Bottom: $V_{\rm NN}<0$. The dashed lines show the actual trace of the parameter space when $\theta$ is varied from $0$ to $\pi/2$. }
\label{fig:ultrav2}
\end{figure}
\subsection*{Ordering of the ground state}
Since the Hamiltonian \noeq{eq:main} without the hopping term is classical, it is fundamentally not difficult to completely identify the lowest energy configuration (see Appendix). The ground states are classified into these three phases: anti-ferromagnetic, dimer and ferromagnetic, depending on the parameters $\delta$, $V_{\rm NN}$ and $V_2$. To explicitly write down the states, $\ket{\text{AFM}}=\ket{\dots \uparrow\downarrow \uparrow\downarrow \dots}$, $\ket{\text{dimer}}=\ket{\dots \uparrow \uparrow \downarrow \downarrow \uparrow \uparrow \dots} $, and $\ket{\text{FM}}=\ket{\dots \uparrow \uparrow \uparrow \downarrow \downarrow \downarrow \dots} $. To ensure half-filling, the ferromagnetic order exhibits a domain-wall in the middle of the system, $\ket{GS} = \ket{\dots \uparrow\uparrow\downarrow\downarrow\dots}$ corresponding to a domain wall soliton \cite{Mikeska2004}. 
We can derive the condition for the system in each of the phases by comparing the energy per site. This is a straightforward task and the result is as follows: anti-ferromagnetic: $E_{GS}/L=V_2/2$, ferromagnetic: $V_2/2+V_{\rm NN}/2$, dimer: $V_{\rm NN}(1-\delta)/4$.  When $V_{\rm NN}<0$, $V_2$ is not relevant and the transition point still lies at $\delta=1$. For the case $\delta >1$ the system is in the dimer phase, and for $\delta<1$, it is in the ferromagnetic phase. When $V_{\rm NN}>0$, $V_2$ significantly affects the phase. When $V_2/V_{\rm NN}<(1-\delta)/2$, the system is in the anti-ferromagnetic phase and when $V_2/V_{\rm NN}<(1-\delta)/2$, it is in the dimer phase. The phase diagram that summarizes the argument is shown in \fig{fig:ultrav2}. Note that interactions and $\delta$ cannot be tuned completely independently. The possible traces are indicated by the gray dashed lines in \fig{fig:ultrav2} with $\gamma = 2\pi/3$ and $5\pi/6$ and $\theta$ varied from 0 to $\pi/2$. It suggests for $\gamma = 2\pi/3$ only one phase transition whereas for $\gamma = 5\pi/6$ there would be two. This can be checked by calculating the derivative with respect to $\theta$ or observing the kinks in the $J_1=0$-curve of \fig{fig:sum2}.

To finish the discussion of the strong coupling limit, we remark that for an odd number of particles the nature of the anti-ferromagnetic and ferromagnetic phases are not altered and merely the ground-state energy will be different. However, in the dimer phase it is easy to see that the additional particle will tend to localize at the edge of the system with a smaller bond-energy. Hence the bulk-state will still show the dimerized structure. 

Before concluding , we would like to mention the case of a small but finite $J_1$ contribution. From the results of the exact diagonalization we see that the cusp at $\theta \sim 0.09\pi$ for $\gamma = 5\pi/6$ (cf. \fig{fig:sum2} graph C), corresponding to the boundary between the anti-ferromagnetic and the dimer-configuration vanishes as soon as $J_1\neq 0$ turning into a smooth crossover. This can be understood intuitively by observing that both states break translational invariance but exhibit a discrete $Z_2$ symmetry, thus belonging to the same symmetry class. On the other hand, the ferromagnetic phase preserves translational invariance and belongs to a different symmetry class. Hence the dimer- and ferromagnetic states cannot be related by a continuous distortion and the cusp remains, as can be seen in \fig{fig:sum2}. Moreover, the numerical results suggest that the phase transition stays of first order even for finite $J_1$ until it vanishes in the TLL phase (see the next section). The transition point is continuously shifted towards large values of $\theta$ with increasing $J_1$. However, the question of whether the first order line and the BKT line meet, and how they close is beyond the scope of this manuscript.

\section{Small coupling case }

Now we will derive a qualitative ground-state phase diagram of this model in the opposite limit -- the case of small dipolar coupling. We assume a finite hopping term $J_1$ and regard the dipolar interaction as a small perturbation, using field-theoretic arguments and a bosonization formalism. In this section, we take the large size ($L \to \infty$) and continuum (lattice spacing $a \to 0$) limits.
The discussion below is a well-studied topic that can be found in standard textbooks in this literature (see for example \cite{Giamarchi2004}) which we closely followed. 

\subsection*{Low energy effective theory of non-interacting fermions}
Rewriting the system in the low energy effective form and in the spin picture,  the non-interacting Hamiltonian becomes
\begin{align}
	H_{XX} &= \sum_j \left[-J_1 (\mS^+_j \mS^-_j +\mS^-_j \mS^+_j)\right] \nonumber \\
	&= \sum_j \left[-J_1(\ad_j \a_{j+1} +\ad_{j+1} \a_j)  \right]  \nonumber \\
	&= -J_1 \int_{-\pi/a}^{\pi/a} dk \cos (ak) \, \tilde{a_k}^\dag \, \tilde{a_k}
	\label{eq:XX}
\end{align}
where in the third line we went into Fourier space. For the case of half-filling, the Fermi points are at $k=\pm\pi/2a$. In the low energy regime, we can linearize the energy spectrum around these Fermi points and introduce slow varying fields. The ground state of this model is now gapless and can be treated as Tomonaga-Luttinger liquid (TLL). Mapping the XX model into an effective low energy model is a well-studied subject and here we will only summarizedthe basic relations to clarify the notations used in this paper.

The Fermi operators can be written as field operators
\begin{equation}
\frac{a_j}{\sqrt{a}}=e^{ik_Fx}\psi_R(x)+e^{-ik_Fx}\psi_L(x)
\end{equation}
where, $k_F = \pi/2a$ and the index $j$ and $x$ are related as $x=ja$. $\psi_R(x)$ and $\psi_L(x)$ are (slow varying) right and left mover operators.  These operators can be described using Boson fields $\varphi_L(x)$ and $\varphi_R(x)$:
\begin{align}
\psi_R(x)=\frac{e^{i\varphi_R(x)}}{\sqrt{2\pi\alpha}} \label{eq:psiR} \\
\psi_L(x)=\frac{e^{-i\varphi_L(x)}}{\sqrt{2\pi\alpha}} \label{eq:psiL}
\end{align}
The $\alpha$ appearing here is an undetermined regularization parameter that has the dimension of length. This mapping from Fermi operator to Boson field operator is called bosonization. While 1D Fermi systems in general show various peculiarities that would make perturbative calculation difficult, the mapped Boson system may be easier to treat. Thus bosonization generally is a effective method in 1D systems.

 It is customary to define the (bosonic) field operators as 
\begin{align}
\phi(x)&=\frac{1}{\sqrt{4\pi}}[\varphi_L(x)+\varphi_R(x)], \nonumber \\
\Pi(x)&=\frac{d}{dx}\frac{1}{\sqrt{4\pi}}[\varphi_L(x)-\varphi_R(x)].
\end{align}
These operators are conjugate and obey the commutation relation $[\phi(x), \Pi(x')]=\delta(x-x')$. Thus, the effective Hamiltonian of the non-interacting spinless fermions (or hard core Bosons) is expressed as
\begin{equation}
\tilde{H}_{XX}=\frac{aJ_1}{2}\int_{-\infty}^{\infty}dx\left[\left(\frac{d\phi}{dx}\right)^2+\Pi^2\right]
\label{eq:BosonizedXX}
\end{equation}
 In order to obtain this form we removed the minus sign that should appear in front of $J_1$.  This can be done in this case as long as the hopping range is NN only:  we (passively) transformed the system by the commutation-conserving transformation of sppin operators  $\mS^x_j \rightarrow \mtS^x_j = (-1)^j \mS^x_j,  \mS^y_j \rightarrow \mtS^y_j=(-1)^j \mS^y_j$, $\mS^z_j \rightarrow \mtS^z_j= \mS^z_j$.  This point is important when interpreting the results of the phase of the system in this effective theory arguments. The tilde mark on the spins and Hamiltonian indicates the transformed expression. 

\subsection*{Bosonization of Ising coupling terms}
Now, the dipolar interaction terms can be added. In spin language, this is simple Ising coupling, the $S^z$ can be written using the Bose field $\phi(x)$ as
\begin{align}
\mtS_j^z&= a_j^\dag a_j-\frac{1}{2} \nonumber \\
&=\frac{a}{\sqrt{\pi}}\frac{d\phi(x)}{dx}+\frac{a(-1)^j}{\pi \alpha}:\sin{\sqrt{4\pi}\phi(x)}: \label{eq:sz}
\end{align}
 where $:\ldots :$ denotes normal ordering.
The nearest neighbor interaction is expressed, by expanding in $a$, as {
\begin{align}
\sum_j \mtS_j^z \mtS_{j+1}^z = a \int_{-\infty}^{\infty} dx \left[ \frac{1}{\pi}\left( \frac{d\phi}{dx} \right)^2+\frac{1}{2\pi^2 \alpha^2}:\cos{(\sqrt{16\pi}\phi)}:+ ...\right]
\label{eq:BosonizedZ}
\end{align}
where $...$ donotes the terms we ignore, which includes quadratic or higher order terms in $a$  and less relevant terms in the context of renormalization group argument such as  $:\cos^2{(\sqrt{16\pi}\phi)}:$ (this point will be explained later).  

The dimerization part of the nearest neighbor interaction ($V_{\rm NN}\,\delta(-1)^j$) requires a different bosonization calculation, due to its oscillatory nature that can lead to back-scattering of a single particle. \cite{Giamarchi2004, Orignac1998}. Expanding in $a$, the bosonized form is expressed as
\begin{align}
\sum_j (-1)^j \mtS_j^z \mtS_{j+1}^z = \frac{a}{\pi \alpha}\int_{-\infty}^{\infty} dx  :\cos{(\sqrt{4\pi}\phi}):+... \label{eq:BosonizedDelta}
\end{align}
Similarly, the next NN interaction term can be written as
\begin{align}
	\sum_j \mtS_j^z \mtS_{j+2}^z = &a\int_{-\infty}^{\infty} \left[-\frac{3}{\pi} \left(\frac{d\phi}{dx}\right)^2 -\frac{1}{2\pi^2 \alpha^2} :\cos(\sqrt{16\pi} \phi: \right] \nonumber \\
	& \, \,+... \label{eq:BosonizedNNN}
\end{align}
Thus,  the form of the zig-zag Hamiltonian (density) is expressed as 
\begin{align}
	\tilde{\mathcal{H}}_{\rm zig-zag} &=  \frac{aJ_1}{2} \left[ \left(1+\frac{4V_{\rm NN}}{\pi J_1}-\frac{6V_2}{\pi J_1} \right) \left(\frac{d\phi(x)}{dx} \right)^2 + \Pi(x)^2 \right] \nonumber \\
	& \quad +\frac{a}{2\pi^2 \alpha^2}  (V_{\rm NN}-V_2):\cos{(\sqrt{16 \pi}\phi)}: \nonumber \\
	& \quad + \frac{\delta V_{\rm NN}}{\pi \alpha} :\cos{(\sqrt{4\pi}\phi)}: \nonumber \\
	&=\frac{u}{2} \left(\frac{1}{K}\left(\frac{d\phi}{dx}\right)^2+K\Pi^2 \right) \nonumber \\
	& \quad \quad + g_1:\cos{(\sqrt{16\pi} \phi)}:+g_\delta:\cos{(\sqrt{4\pi}\phi)}:
	\label{eq:fullform} 
\end{align}
where again we ignored higher order terms in $a$, and operators with higher oscillation frequencies that are less relevant in terms of following renormalization group argument. The $K$ and $u$ are the Luttinger parameters, calculated to be
\begin{align}
K&=\frac{1}{\sqrt{1+\frac{4\Delta_1-6\Delta_2}{\pi}}} ,\nonumber \\
u&= aJ_1\sqrt{1+\frac{4\Delta_1-6\Delta_2}{\pi}} ,\nonumber \\ \label{eq:constants}
\end{align}
where $\Delta_1=V_{\rm NN}/J_1, \Delta_2=V_2/J_1$. The result is accurate up to first order in $\Delta_1$ and $\Delta_2$.  The $g_1, g_\delta$ are non-universal coupling constants. This ``non-universality'' stems from the remaining cut-off parameters $a$ and $\alpha$ appearing in these constants. In order to accurately determine these constants one would need to take into account all orders of the expansion in \eqs{eq:BosonizedZ} -- \noeq{eq:BosonizedNNN}. In most cases, this is impossible analytically. This solution, however, gives a good qualitative picture of the system.

 We observe that, as the angle of molecules $\theta$ changes, $\Delta_1$ and $\Delta_2$ dramatically change, and consequently the Luttinger parameter $K$ can take a wide range of values, resulting in a rich phase diagram. $K$ determines the asymptotic behavior of the system's correlation function in TLL, such as the charge-density wave (CDW) correlation function $c_{\rm CDW}$. Working at zero magnetic field this is given by
\begin{align}
	c_{\rm CDW} \propto \left\langle\mtS^z(x)\mtS^z(0)\right\rangle \sim\frac{K}{2\pi}\frac{1}{x^2}+ A\cos(2 \pi \rho_0 x)\frac{1}{x^{2K}}, 
	\label{eq:CDW_correlation}
\end{align}
with a non-universal amplitude $A$ and $\rho_0 = 1/(2a)$.

\begin{figure*}
\includegraphics[scale=0.5]{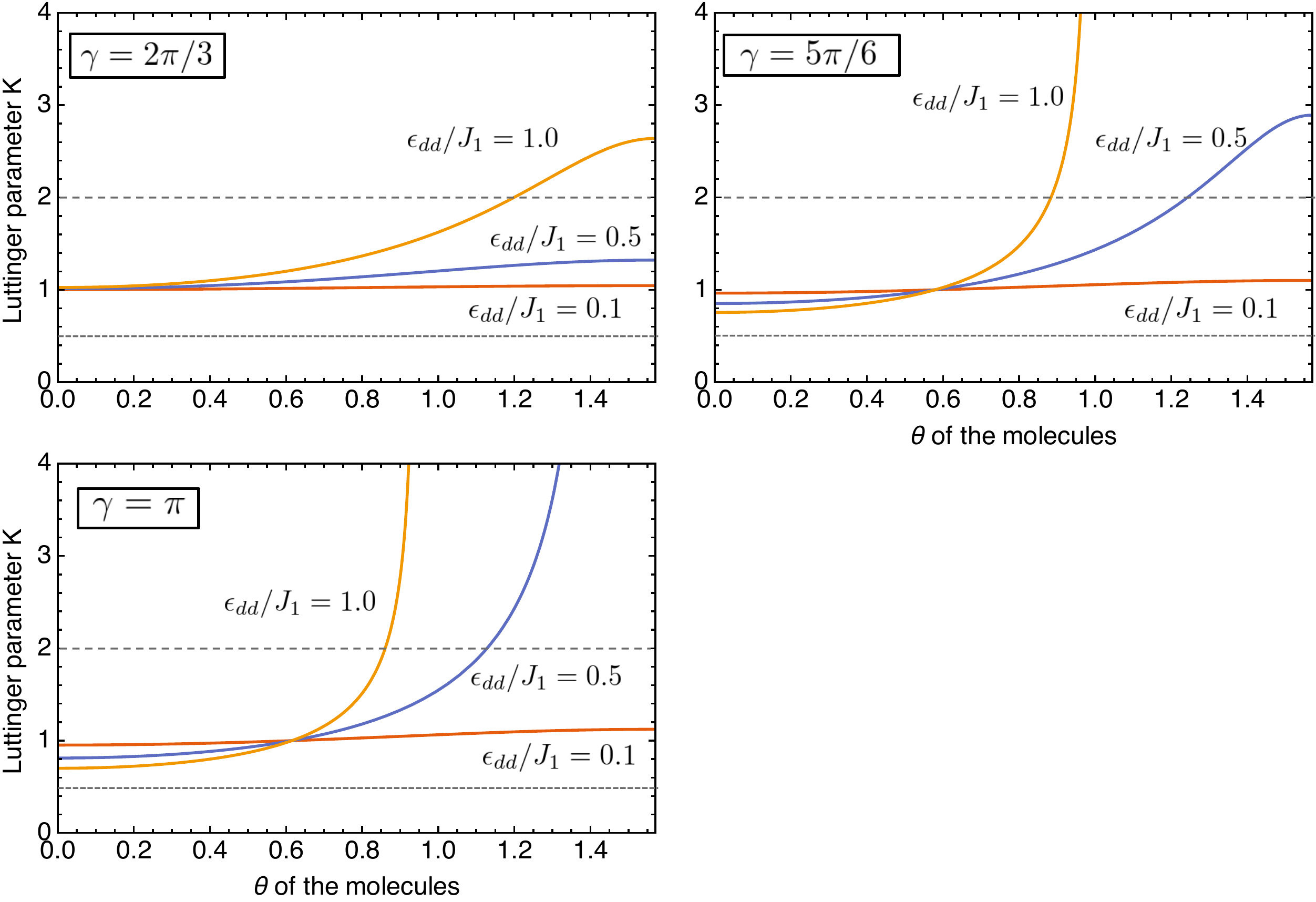}
\caption{\label{fig:lutK} (Color online) Luttinger parameter plotted vs. $\theta$ of the molecules, for different hopping parameters $J_1$ and for different zig-zag angles $\gamma$. These results are based on \eq{eq:constants}, calculated using perturbative renormalization group arguments. }
\end{figure*}

\subsection*{Renormalization Group Arguments}
The ordering of the system is qualitatively discussed using a first order renormalization group argument of the Sine-Gordon Hamiltonian \noeq{eq:fullform}. The renormalization group argument of this model is well known \cite{Giamarchi2004} and here we apply the result. First we investigate the relevance of the non-linear terms $g_1 \cos{(\sqrt{16\pi}\phi)}$ and $g_\delta \cos{(\sqrt{4\pi}\phi)}$. In general, the scaling dimension of an operator of $g \exp{(i\sqrt{4n^2\pi}\phi)}$ type is known to be $n^2K$, where $K$ is the usual Luttinger parameter and $g$ is the coupling constant. The scaling equation is known to be \cite{Giamarchi2004}
\begin{align}
\frac{dg}{dl}=(2-n^2K)g, \quad \frac{dK}{dl}=-Cg^2a^4,
\end{align}
implying that for $K = 2/n^2$ the $g \exp{(i\sqrt{4n^2\pi}\phi)}$ operator is marginal, while it is irrelevant for smaller values of $K$.
In the case of our Hamiltonian, $n=2$ for the $g_1 :\cos{(\sqrt{16\pi} \phi)}:$ term implying this operator changes its relevance at $K=1/2$, and similarly $n=1$ for the $g_\delta :\cos{(\sqrt{4\pi} \phi)}:$ term, changing its relevance at $K=2$ . 

When $K<1/2$,  the $g_1 :\cos{(\sqrt{16\pi}\phi)}:$ term becomes relevant and the system is entirely governed by this term. The system in this case is driven to either anti-ferromagnetic or dimer order, depending on the sign of the coupling constant $g_1$. When $g_1>0$, the bosonic field $\phi(x)$ appearing in the $g_1 :\cos{(\sqrt{16\pi}\phi)}:$ tries to minimize this term and takes the value such that $\sqrt{16\pi}\phi(x)=\pi$ or $\phi(x)=\sqrt{\pi}/4$. From \eq{eq:sz} we see that $\mtS^z \approx (-1)^j \sin{(\pi/2+n\pi)} \approx (-1)^j$, i.e., spin changes its sign at every each site. On the other hand if $g_1<0$, the bosonic field is pinned to $\phi=0$, leading  to $\left\langle\mtS^z_j\right\rangle=0$ and the finite dimer value of  $\vec{S}_i \cdot \vec{S}_{i+1}-\vec{S}_{i+1} \cdot\vec{S}_{i+2} =(-1)^j$. As before, perturbative theory cannot in general determine the sign of $g_1$, thus the differentiation between dimer and anti-ferromagnetic phases has to be done numerically.  We will see (cf. \fig{fig:lutK}), however, that in our system for large $J_1$, $K$ is always larger than $1/2$, implying $g_1 :\cos{(\sqrt{16\pi}\phi)}:$ is always irrelevant, and thus we do not go further to discuss this point.

  When $K<2$, the term $g_\delta :\cos{(\sqrt{4\pi}\phi)}:$ is relevant.   For our system, we see that $K<2$ is satisfied in a broad region (above dashed line in \fig{fig:lutK}). In particular, this is true even if the system is barely interacting, namely when $\Delta_1$ and $\Delta_2$ are both close to 0. This implies that the $g_\delta \cos{(\sqrt{4\pi}\phi)}$ term is relevant and the system is governed by this term no matter how small the interaction and dimerization are, as long as they remain finite.  This behavior has been described as ``Spin-Peierls instability'' \cite{Cross1979,Orignac2004} -- even a tiny distortion of the lattice (in our case the dimerization) will open an energy gap. The gap scales as $E_g \propto \delta^{1/(2-K)}$. In a case like this, the bosonic field $\phi$ tries to minimize $g_\delta :\cos{(\sqrt{4\pi}\phi)}:$. As a result, $\sqrt{4\pi}\phi=\pm\pi$, depending on the sign of $g_\delta$, and thus $\left\langle\mtS^z_j\right\rangle=0$ (cf.\eq{eq:sz}),  and  $\vec{S}_i \cdot \vec{S}_{i+1}-\vec{S}_{i+1} \cdot\vec{S}_{i+2} = (-1)^j$, i.e., resulting in dimerized order. Unlike the previous case, the sign of $g_\delta$ does not qualitatively change the order. Although the coupling inducing the dimerization is Ising-like, for large $J_1 \gg \epsilon_{dd}$ this dimerized state is a {\it valence bond  state} (VBS), which is explicitly expressed as 
$(\ket{\uparrow\downarrow}-\ket{\downarrow \uparrow}/\sqrt{2}) \otimes (\ket{\uparrow\downarrow}-\ket{\downarrow \uparrow}/\sqrt{2}) \otimes...  $, in contrast to the Ising type dimer state ($\ket{\text{dimer}}=\ket{\dots \uparrow \uparrow \downarrow \downarrow \uparrow \uparrow \dots} $) that appeared in previous sections. 

The case of $K<0$ in \eq{eq:fullform} cannot be discussed in the same manner because minimizing the coupling terms $g_1 :\cos{(\sqrt{16\pi}\phi)}:$ and $g_\delta :\cos{(\sqrt{4\pi}\phi)}:$ does not give insight into the phase diagram. Yet, \eq{eq:constants} indicates that this case appears for both strong attractive NN interaction and repulsive NNN interaction, implying that the system will be in the dimerized or ferromagnetic phase depending on the parameters $\theta, \gamma$ and $J_1$, but not in the anti-ferromagnetic state.

Going beyond perturbation in $\Delta_1, \Delta_2$, the Luttinger parameter $K$ has to be found numerically. There are, however, 
special points in the parameter space where  $K$ and $u$ can be obtained analytically. For example, for $\Delta_2=0, \delta=0$ the model reduces to the XXZ model which allows for an exact calculation, using e.g. Bethe-Ansatz techniques \cite{Sirker2012}. 

\begin{align}
	&K=\frac{1}{2(1-\pi^{-1}\cos{^{-1}(\Delta_1)})} \nonumber \\
	&u=\frac{\pi\sqrt{1-\Delta_1^2} }{2\cos{^{-1}\Delta_1}}
	\label{eq:bethe}
\end{align}
and thus $K\in [1/2, \infty)$. To check, it can be seen that K in \eq{eq:bethe} has the same form up to first order in $\Delta_1$ as \eq{eq:constants}
\begin{figure*}
\includegraphics[width=\linewidth]{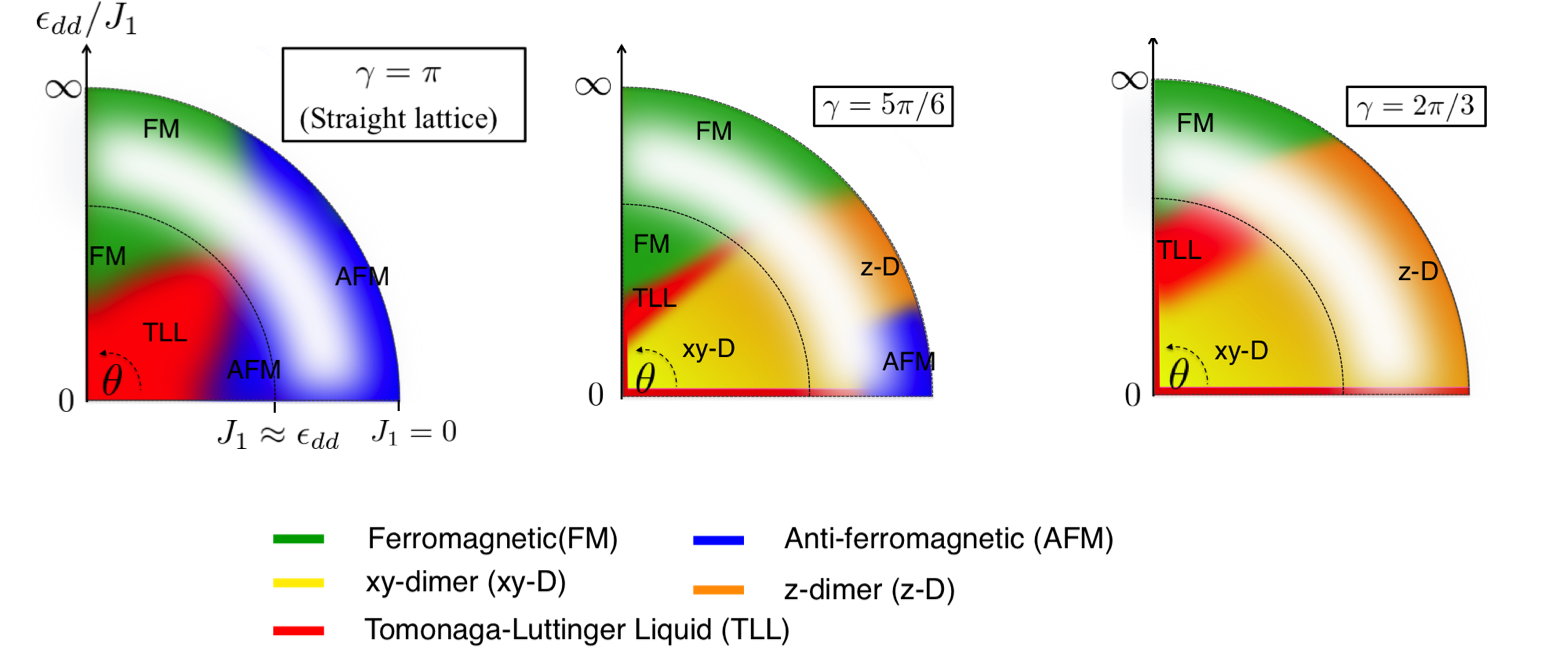}
\caption{ (Color online) Qualitative ground state phase diagram of our system. The radial degree of freedom shows the inverse of the hopping parameter of the system, and the argument $\theta$ is the angle of the polarized molecules. Each color shows a different phase. The white shaded area is the region whose ordering behavior is not studied in this paper.  }
\label{fig:phasediagram}
\end{figure*}

\subsection*{Phase Diagrams}

The phase diagrams that sum up the discussion are shown in \fig{fig:phasediagram}.
The phase diagrams are the result of field theoretical analysis using bosonization techniques and first order perturbative renormalization group arguments. They inevitably involve approximations and therefore unspecified constants, resulting in an overall qualitative picture of the system rather than a quantitative one.  At this point, numerical methods are needed to accurately determine many exact transitions in the phase diagram. The qualitative discussion so far, however, provides a good picture of the overall behavior of the system. Each pie in \fig{fig:phasediagram} shows the ground state phases for three different zig-zag angles $\gamma$, where the first, $\gamma=\pi$, is just the solution of a straight chain. The diagrams are depicted in polar coordinates, showing  the ratio of lattice depth to hopping as the radius and he angle argument as the actual polarization angle $\theta$ of the molecules. 

The border between the TLL phase and other phases indicates the Berezinsky-Kosterlitz-Thouless (BKT) transition. When $J_1$ is  large, the gapped phases  border to the gapless TLL phase , and the system is expected to be dominated by BKT  transitions, as $\theta$ changes from 0 to $\pi/2$. In contrast to that, when $J_1$ is small, the phases are connected with first order transition lines. We have not adequately studied the region around $J_1 \approx 1$, and hence the details of crossing of first order and the BKT lines  are beyond the scope of this paper.

  We here remind ourself that we transformed the spin operators in the beginning of bosonization treatment.  In order to going back to the original (untransformed) system, one simply needs to perform the same spin operator transformation again $\mtS^x_j \rightarrow \mS^x_j = (-1)^j \mtS^x_j,  \mtS^y_j \rightarrow (-1)^j \mS^y_j$, $\mtS^z_j \rightarrow \mS^z_j= \mtS^z_j$.  Therefore the reinterpretation is simply equivalent to acting with a unitary operator $U=U^{-1}=\sigma_z=\ket{\uparrow}\bra{\uparrow}-\ket{\downarrow}\bra{\downarrow}$ on every other lattice site. The phase diagram shows the result in the language of untransformed spins. An important consequence of this remapping is that the VBS state is now remapped into a triplet bound state, or explicitly, $(\ket{\uparrow\downarrow}+\ket{\downarrow \uparrow}/\sqrt{2}) \otimes (\ket{\uparrow\downarrow}+\ket{\downarrow \uparrow}/\sqrt{2}) \otimes...  $. We call this state an ``xy-dimer'' as the dimerized pairs can be seen as polarized in the xy-plane as opposed to the dimerized order for $J_1 \approx 0$ where the dimerized pairs are polarized in the z-direction (we call this type ``z-dimer''). One observes that when the opening angle $\gamma$ is smaller than $\pi$, the system is predominantly in the xy-dimer state when $J_1 \to \infty$ where each dimer in spin language is $\ket{L,M_L}=\ket{1, 0}$. (Here, $L$ is the total spin of the dimerized pair). As the optical lattice deepens (i.e., a move in radial direction in phase diagram) the dimer pair will gradually polarize into the z-direction by picking up the $M_L=1$ component until it becomes completely polarized in the z-direction, becoming $\ket{1,1}$. Thus, the depth of the optical lattice tunes the polarization direction of the dimerized pairs.

\section{Conclusion}

The zigzag nature of the chain induces bond-alternating nearest-neighbor interactions as a function of the molecules' aligned angle with the chain axis. We also have taken up to next nearest neighbor interaction to include the long range nature of the dipole interaction, introducing (Ising-type) frustration in the system. In the strong coupling limit, the ground state ordering is exactly identified, where the system lies in either anti-ferromagnetic, ferromagnetic or Ising-dimer, depending on the coupling parameters. In the weak coupling limit, the effective field theory additionally predicts TLL phase and dimerized phase, whose dimerized pairs have different polarized direction than the strong coupling case. The polarization of dimerized pairs should be closely affected by the depth of the optical lattice. Our methods do not accurately predicts the ordering in the region of intermediate hopping and this should be the included in the future works. 

The goal is to utilize this simple quasi-1D model to see phases beyond typical 1D physics. While we here discussed only the phase-diagram of polarized hard-core dipoles at half-filling, moving on a 1D zig-zag chain, first, the richness of the system is obvious in the phase diagrams shown above. Second, the extension to other filling ratios and not only longer-range interactions but also longer-range hopping is obvious and very experimentally feasible. This should lead to very interesting quantum fluctuation that can lead to unconventional quantum phases \cite{Furukawa2012a,Furukawa2010b}. Exploring similar phases with smaller $\gamma$ in our model will be subject to future studies.  Moving away from half-filling and taking longer-range parts of the dipolar interaction into account can lead to interesting modification of the Devils's staircase \cite{Burnell2009}.



\color{black}
\subsection*{Acknowledgements}
We are grateful for many insightful discussions with J. F. Kong, A. Shashi, E. Dalla Torre, T. Calarco, and N. Ghimire. QYW is grateful for the financial support from Japan Student Services Organization (JASSO).  JO acknowledges financial support from the Harvard Quantum Optics Center. This project was supported by NSF, ASOFR, and was started at KITP.

\appendix
\section{\\Identifying the ground states when $J_1=0$} \label{App:AppendixA}
 Here we would like to show the outline to obtain the phase diagram in ultrastrong coulpling case.
 Because I argue that either of the 3 phases are the ground state, I would like to explicitly construct a state and compare its energy per site with either of 3 phases. Here I use symbols like \fbox{n L} (or \fbox{n R}) $n=1,2,3...$ to denote a "building block" of the system, whose meaning is $n$ left (right) sites are filled and n right (left) sites are empty.  For example, \fbox{1, L} is $\bullet \circ$ with black circle being the filled sites and white circle empty sites. \fbox{3, R} is $\circ \circ \circ \bullet \bullet \bullet$ and so on. 

Using these "blcok" notations, the 3 presumable ground states, ferromagnetic, anti-ferromagnetic and dimer state are described as 
\begin{align}
 \ket{AFM} &= \fbox{1, L}-\fbox{1, L}-\fbox{1, L}-\fbox{1, L}- .... \nonumber \\
 \ket{FM}&= \fbox{N/2, L} \nonumber \\
 \ket{Dimer}&= \fbox{2, L}-\fbox{2, L}-\fbox{2, L}-\fbox{2, L}- ... 
\end{align}
 and their average energy per site is 
 \begin{align}
  E_{p.s.}^{AFM} &= \frac{V_2}{2} \nonumber \\
 E_{p.s.}^{FM} &= \frac{V_1^{even}+V_1^{odd}+2V_2}{4} \nonumber \\
 E_{p.s.}^{Dimer} &= \frac{V_1^{even}}{4}
\end{align}

 It is useful to investigate the energy density of these "building blocks" for the later comparison. It is easy to convince oneself that the average energy "per site" is different depending on whether $N$ is odd or even. To write it explicitly with the coupling constants $V_1^{odd}, V_1^{even}, V_2$,
 \begin{align}E_{p.s.}(\fbox{2m, L}) &=\frac{mV_1^{even}+(m-1)V_1^{odd}+(2m-2)V_2}{4m} \nonumber \\
  E_{p.s.}(\fbox{2m+1, L}) &= \frac{mV_1^{even}+mV_1^{odd}+(2m-1)V_2}{4m+2} 
\end{align}

The difference of these energy are computed as,
 \begin{align}\Delta E_{p.s.}^{2m L} & \equiv E_{p.s.}(\fbox{2(m+1), L})-E_{p.s.}(\fbox{2m, L}) \nonumber \\ 
 &=\frac{2V_1^{odd}+V_2}{2m(m+1)} \nonumber \\
  \Delta E_{p.s.}^{2m+1 L} & \equiv E_{p.s.}(\fbox{2(m+1)+1, L})-E_{p.s.}(\fbox{2m+1, L}) \nonumber \\
  &=\frac{V_1^{odd}+V_1^{even}+4V_2}{2(2m+3)(2m+1)} \label{eq:difference}
\end{align}
We see that depending on the sign and magnitude of the interaction parameters, $\Delta E_{p.s.}$ can be positive or negative (or 0), regardless of the value of $m$ (or equivalently $N$). This means that when we fix $V_1^{odd}, V_1^{even}, V_2$ the energy per site of the building blocks are either monotonic increase or decrease or just a constant with respect to $m$ and thus we can find a unique building block that has the lowest energy per site. We can presume that the ground state is build with these lowest energy per site building blocks. 

However we need to take into account the "connection energy" arising from additional interaction between the connecting building blocks. For example \fbox{1, L} -- \fbox{N, L} with $N \geq 2$ ( $\bullet \circ$ -- $\bullet \bullet \bullet ... \circ \circ \circ $  ) generates $V_2$ upon connecting (Remember $\bullet$ is filled and $\circ$ is empty site.)  Since the range of the interaction is at most 2 sites, the contribution of the connection is the same for all $N \geq 2$. Therefore when we consider the connection, it is sufficient to classify the building blocks into 4 cases: \fbox{1, L}, \fbox{1, R}, \fbox{N, L}, \fbox{N, R}, with $N\geq 2$. To list up all possible connections, there are $4 \times 4 = 16$ possible possibilities -- one of those 4 building blocks on the left and one of those 4 on the right. All connections are shown in Table (\ref{table:connection})

Before moving onto the next step, we note that the role of $V_1^{odd}$ and $ V_1^{even}$ can be flipped by inserting an empty site at the left edge of the chain.  Instead of performing this we remove this redundancy by deliberatively force $V_1^{odd} \geq V_1^{even}$ or vise versa, depending on in each case. For example, $\ket{Dimer}$ has average energy $V_1^{even}/4$ per site. By inserting an additional site (or translating by 1 site) the energy is $V_1^{odd}/4$. In this situation we will just assume $V_1^{odd}<V_1^{even}$.
With all of these information, we would like to explicitly construct a state that has the lowest energy with given interaction constants and prove that the either one of the 3 phases (dimer, ferromagnetic, anti-ferromagnetic) has the lowest energy in any case. From now on, we use $n$ to be general integer that is larger than or equal to 0, and $N$ to be integer that is larger than or equal to 2. 

\begin{table}
	\caption[Connection energy]{the building blocks on the left are the left component of the connection. The top ones are right component of the connection. For example (3,2) element of the table, $V_1^{odd}$, indicates \fbox{1,R}--\fbox{N, L} connection of gives $V_1^{odd}$ energy. }
	\label{table:connection}
	\begin{tabular}{c| c c c c}
		test \, & \, \fbox{1, L} \,  & \fbox{1, R} \,  & \fbox{N, L} \,  & \fbox{N, R} \\ \hline
		\fbox{1, L} \, & $V_2$  & 0 & $V_2$ & 0 \\
		\fbox{1, R} \, & $V_1^{odd}$ & $V_2$ & $V_1^{odd}$ & 0 \\
		\fbox{N, L} \, &0 &0 &0 &0 \\
		\fbox{N, R} \, & $V_1^{odd}+V_2$ & $V_2$ &$V_1^{odd}+2V_2$ &0
	\end{tabular}
\end{table}

When interaction parameters fulfill these conditions the ground state is obvious.
\begin{enumerate}
	\item $V_1^{even}<0, V_1^{odd}<0, V_2<0 \Rightarrow \ket{FM}$
	\item $V_1^{even}>0, V_1^{odd}<0, V_2>0$ or  $V_1^{even}<0, V_1^{odd}>0, V_2>0 \Rightarrow \ket{Dimer}$
	\item $V_1^{even}>0, V_1^{odd}>0, V_2<0 \Rightarrow \ket{AFM}$
\end{enumerate}
Now let us tackle on the less obvious case. We need to consider these 4 cases: 
\begin{enumerate}
  \item $V_1^{even}>0, V_1^{odd}>0, V_2>0$
  \item $V_1^{even}<0, V_1^{odd}<0, V_2>0$
  \item $V_1^{even}>0, V_1^{odd}<0, V_2<0$
  \item $V_1^{even}<0, V_1^{odd}>0, V_2<0$
\end{enumerate}

 First let us look into case 1. ($V_1^{even}, V_1^{odd}, V_2 >0$). For simplicity, we can impose another condition, that is $V_1^{even}<V_1^{odd}$. Then we prove that when $V_2>V_1^{even}/2$ the lowest energy state is Dimer with average energy per site $E_{p.s.}^{Dimer}=V_1^{even}/4$ and when $V_2<V_1^{even}/2$ it is in anti-ferromagentic order and $E_{p.s.}^{AFM}=V_2/2$ just by explicitly computing the energy.

Now consider a general state 
\begin{equation}
\underbrace{\fbox{n', L}-...}_{\text{made of \fbox{L}}}-\underbrace{\fbox{n'', R}-...}_{\text{made of \fbox{R}}}-\fbox{n''', L}-.... \label{eq:general_state}
\end{equation}

When all the interactions are positive, from Eq.(\ref{eq:difference}), we know that the average energy per site of the building blocks is the smallest when $n=1$. This let us exclude the possibility of $n, n', n'' >2$ that appears in Eq.(\ref{eq:general_state}). So the ground state must be built with "building blocks" whose $n$ is either 1 or 2.

 Let's assume all $n$ appears in Eq.(\ref{eq:general_state}) are 1. Looking at the \fig{table:connection}, we see that the possible lowest energy state is either
\fbox{1, L}-\fbox{1, R}-\fbox{1, L}-\fbox{1, R}-\fbox{1, L}.... 
whose average energy per site is $V_1^{odd}/4$,  or
\fbox{1,L}-\fbox{1,L}-\fbox{1, L}-.... (or equivalently \fbox{1, R}-\fbox{1, R}-...) 
whose average energy per site is $V_2/2$. 

Similarly, when we set all $n=2$, possible lowest energy state is \fbox{2,L}--\fbox{2,L}--\fbox{2, L}...or \fbox{2,R}--\fbox{2,R}--\fbox{2,R}-- and the average energy per site is $V_1^{even}/4$. 

From these analysis we set an upper bound for the ground state average energy per site: 
 \begin{align}
 V_1^{even}<V_1^{odd}  \land  V_2<V_1^{even}/2 &\Rightarrow E_{G.S} \leq V_2/2 \nonumber \\
 V_1^{even}<V_1^{odd} \land V_2>V_1^{even}/2 &\Rightarrow E_{G.S} \leq V_1^{even}/4 \nonumber \\ \label{eq:gsbound}
 \end{align}
Now we need to take into account the third case -- state with $n=1$ and $n=2$ "building blocks" combined. One can come up with low energy states such as 
\fbox{2, L}-\fbox{1, R}-\fbox{2, R} - \text{repetition of this set of 3 blocks}, whose energy per site is $\frac{2V_1^{even}+V_1^{odd}+2V_2}{10}$ and \fbox{1, L}-\fbox{1, R} -\fbox{2, L} -\text{repetition of this set of 3 blocks}, whose energy per site is $\frac{V_1^{odd}+V_1^{even}}{8}$. Both of these energy exceeds the upper bound we set previously at Eq.(\ref{eq:gsbound}) and cannot be the ground state. Therefore the ground state configuration must be either \fbox{2,L}--\fbox{2,L}--\fbox{2, L}...or \fbox{2,R}--\fbox{2,R}--\fbox{2,R}--, meaning the ground state is the (Ising) dimer phase. 

The other ground states for less obvious cases can be identified exactly the same way and we will not list the derivation here. Again the results that summarize this section is shown in \fig{fig:ultrav2}.

\bibliographystyle{apsrev4-1}
\bibliography{zigzag_Qing}

\begin{thebibliography}{37}%
\makeatletter
\providecommand \@ifxundefined [1]{%
 \@ifx{#1\undefined}
}%
\providecommand \@ifnum [1]{%
 \ifnum #1\expandafter \@firstoftwo
 \else \expandafter \@secondoftwo
 \fi
}%
\providecommand \@ifx [1]{%
 \ifx #1\expandafter \@firstoftwo
 \else \expandafter \@secondoftwo
 \fi
}%
\providecommand \natexlab [1]{#1}%
\providecommand \enquote  [1]{``#1''}%
\providecommand \bibnamefont  [1]{#1}%
\providecommand \bibfnamefont [1]{#1}%
\providecommand \citenamefont [1]{#1}%
\providecommand \href@noop [0]{\@secondoftwo}%
\providecommand \href [0]{\begingroup \@sanitize@url \@href}%
\providecommand \@href[1]{\@@startlink{#1}\@@href}%
\providecommand \@@href[1]{\endgroup#1\@@endlink}%
\providecommand \@sanitize@url [0]{\catcode `\\12\catcode `\$12\catcode
  `\&12\catcode `\#12\catcode `\^12\catcode `\_12\catcode `\%12\relax}%
\providecommand \@@startlink[1]{}%
\providecommand \@@endlink[0]{}%
\providecommand \url  [0]{\begingroup\@sanitize@url \@url }%
\providecommand \@url [1]{\endgroup\@href {#1}{\urlprefix }}%
\providecommand \urlprefix  [0]{URL }%
\providecommand \Eprint [0]{\href }%
\providecommand \doibase [0]{http://dx.doi.org/}%
\providecommand \selectlanguage [0]{\@gobble}%
\providecommand \bibinfo  [0]{\@secondoftwo}%
\providecommand \bibfield  [0]{\@secondoftwo}%
\providecommand \translation [1]{[#1]}%
\providecommand \BibitemOpen [0]{}%
\providecommand \bibitemStop [0]{}%
\providecommand \bibitemNoStop [0]{.\EOS\space}%
\providecommand \EOS [0]{\spacefactor3000\relax}%
\providecommand \BibitemShut  [1]{\csname bibitem#1\endcsname}%
\let\auto@bib@innerbib\@empty
\bibitem [{\citenamefont {Carr}\ \emph {et~al.}(2009)\citenamefont {Carr},
  \citenamefont {DeMille}, \citenamefont {Krems},\ and\ \citenamefont
  {Ye}}]{Carr2009}%
  \BibitemOpen
  \bibfield  {author} {\bibinfo {author} {\bibfnamefont {L.~D.}\ \bibnamefont
  {Carr}}, \bibinfo {author} {\bibfnamefont {D.}~\bibnamefont {DeMille}},
  \bibinfo {author} {\bibfnamefont {R.~V.}\ \bibnamefont {Krems}}, \ and\
  \bibinfo {author} {\bibfnamefont {J.}~\bibnamefont {Ye}},\ }\href {\doibase
  10.1088/1367-2630/11/5/055049} {\bibfield  {journal} {\bibinfo  {journal}
  {New Journal of Physics}\ }\textbf {\bibinfo {volume} {11}} (\bibinfo {year}
  {2009}),\ 10.1088/1367-2630/11/5/055049},\ \Eprint
  {http://arxiv.org/abs/0904.3175} {arXiv:0904.3175} \BibitemShut {NoStop}%
\bibitem [{\citenamefont {Baranov}\ \emph {et~al.}(2012)\citenamefont
  {Baranov}, \citenamefont {Dalmonte}, \citenamefont {Pupillo},\ and\
  \citenamefont {Zoller}}]{Baranov2012}%
  \BibitemOpen
  \bibfield  {author} {\bibinfo {author} {\bibfnamefont {M.~A.}\ \bibnamefont
  {Baranov}}, \bibinfo {author} {\bibfnamefont {M.}~\bibnamefont {Dalmonte}},
  \bibinfo {author} {\bibfnamefont {G.}~\bibnamefont {Pupillo}}, \ and\
  \bibinfo {author} {\bibfnamefont {P.}~\bibnamefont {Zoller}},\ }\href
  {\doibase 10.1021/cr2003568} {\bibfield  {journal} {\bibinfo  {journal}
  {{Chem. Rev.}}\ }\textbf {\bibinfo {volume} {112}},\ \bibinfo {pages} {5012}
  (\bibinfo {year} {2012})}\BibitemShut {NoStop}%
\bibitem [{\citenamefont {Becker}\ \emph {et~al.}(2010)\citenamefont {Becker},
  \citenamefont {Soltan-Panahi}, \citenamefont {Kronj\"{a}ger}, \citenamefont
  {D\"{o}rscher}, \citenamefont {Bongs},\ and\ \citenamefont
  {Sengstock}}]{Becker2010}%
  \BibitemOpen
  \bibfield  {author} {\bibinfo {author} {\bibfnamefont {C.}~\bibnamefont
  {Becker}}, \bibinfo {author} {\bibfnamefont {P.}~\bibnamefont
  {Soltan-Panahi}}, \bibinfo {author} {\bibfnamefont {J.}~\bibnamefont
  {Kronj\"{a}ger}}, \bibinfo {author} {\bibfnamefont {S.}~\bibnamefont
  {D\"{o}rscher}}, \bibinfo {author} {\bibfnamefont {K.}~\bibnamefont {Bongs}},
  \ and\ \bibinfo {author} {\bibfnamefont {K.}~\bibnamefont {Sengstock}},\
  }\href {\doibase 10.1088/1367-2630/12/6/065025} {\bibfield  {journal}
  {\bibinfo  {journal} {{New J. Phys.}}\ }\textbf {\bibinfo {volume} {12}},\
  \bibinfo {pages} {065025} (\bibinfo {year} {2010})}\BibitemShut {NoStop}%
\bibitem [{\citenamefont {Windpassinger}\ and\ \citenamefont
  {Sengstock}(2013)}]{Windpassinger2013}%
  \BibitemOpen
  \bibfield  {author} {\bibinfo {author} {\bibfnamefont {P.}~\bibnamefont
  {Windpassinger}}\ and\ \bibinfo {author} {\bibfnamefont {K.}~\bibnamefont
  {Sengstock}},\ }\href {\doibase 10.1088/0034-4885/76/8/086401} {\bibfield
  {journal} {\bibinfo  {journal} {{Rep. Prog. Phys.}}\ }\textbf {\bibinfo
  {volume} {76}},\ \bibinfo {pages} {086401} (\bibinfo {year}
  {2013})}\BibitemShut {NoStop}%
\bibitem [{\citenamefont {Nogrette}\ \emph {et~al.}(2014)\citenamefont
  {Nogrette}, \citenamefont {Labuhn}, \citenamefont {Ravets}, \citenamefont
  {Barredo}, \citenamefont {B\'{e}guin}, \citenamefont {Vernier}, \citenamefont
  {Lahaye},\ and\ \citenamefont {Browaeys}}]{Nogrette2014a}%
  \BibitemOpen
  \bibfield  {author} {\bibinfo {author} {\bibfnamefont {F.}~\bibnamefont
  {Nogrette}}, \bibinfo {author} {\bibfnamefont {H.}~\bibnamefont {Labuhn}},
  \bibinfo {author} {\bibfnamefont {S.}~\bibnamefont {Ravets}}, \bibinfo
  {author} {\bibfnamefont {D.}~\bibnamefont {Barredo}}, \bibinfo {author}
  {\bibfnamefont {L.}~\bibnamefont {B\'{e}guin}}, \bibinfo {author}
  {\bibfnamefont {a.}~\bibnamefont {Vernier}}, \bibinfo {author} {\bibfnamefont
  {T.}~\bibnamefont {Lahaye}}, \ and\ \bibinfo {author} {\bibfnamefont
  {a.}~\bibnamefont {Browaeys}},\ }\href {\doibase 10.1103/PhysRevX.4.021034}
  {\bibfield  {journal} {\bibinfo  {journal} {Physical Review X}\ }\textbf
  {\bibinfo {volume} {4}},\ \bibinfo {pages} {1} (\bibinfo {year} {2014})},\
  \Eprint {http://arxiv.org/abs/1402.5329} {arXiv:1402.5329} \BibitemShut
  {NoStop}%
\bibitem [{\citenamefont {Ohgoe}\ \emph {et~al.}(2012)\citenamefont {Ohgoe},
  \citenamefont {Suzuki},\ and\ \citenamefont {Kawashima}}]{Ohgoe2012}%
  \BibitemOpen
  \bibfield  {author} {\bibinfo {author} {\bibfnamefont {T.}~\bibnamefont
  {Ohgoe}}, \bibinfo {author} {\bibfnamefont {T.}~\bibnamefont {Suzuki}}, \
  and\ \bibinfo {author} {\bibfnamefont {N.}~\bibnamefont {Kawashima}},\ }\href
  {\doibase 10.1103/PhysRevA.86.063635} {\bibfield  {journal} {\bibinfo
  {journal} {{Phys. Rev. A}}\ }\textbf {\bibinfo {volume} {86}},\ \bibinfo
  {pages} {063635} (\bibinfo {year} {2012})}\BibitemShut {NoStop}%
\bibitem [{\citenamefont {Albuquerque}\ \emph {et~al.}(2011)\citenamefont
  {Albuquerque}, \citenamefont {Schwandt}, \citenamefont {Het\'{e}nyi},
  \citenamefont {Capponi}, \citenamefont {Mambrini},\ and\ \citenamefont
  {L\"{a}uchli}}]{Albuquerque2011}%
  \BibitemOpen
  \bibfield  {author} {\bibinfo {author} {\bibfnamefont {A.~F.}\ \bibnamefont
  {Albuquerque}}, \bibinfo {author} {\bibfnamefont {D.}~\bibnamefont
  {Schwandt}}, \bibinfo {author} {\bibfnamefont {B.}~\bibnamefont
  {Het\'{e}nyi}}, \bibinfo {author} {\bibfnamefont {S.}~\bibnamefont
  {Capponi}}, \bibinfo {author} {\bibfnamefont {M.}~\bibnamefont {Mambrini}}, \
  and\ \bibinfo {author} {\bibfnamefont {A.~M.}\ \bibnamefont {L\"{a}uchli}},\
  }\href {\doibase 10.1103/PhysRevB.84.024406} {\bibfield  {journal} {\bibinfo
  {journal} {{Phys. Rev. B}}\ }\textbf {\bibinfo {volume} {84}},\ \bibinfo
  {pages} {024406} (\bibinfo {year} {2011})}\BibitemShut {NoStop}%
\bibitem [{\citenamefont {Mezzacapo}\ and\ \citenamefont
  {Boninsegni}(2012)}]{Mezzacapo2012}%
  \BibitemOpen
  \bibfield  {author} {\bibinfo {author} {\bibfnamefont {F.}~\bibnamefont
  {Mezzacapo}}\ and\ \bibinfo {author} {\bibfnamefont {M.}~\bibnamefont
  {Boninsegni}},\ }\href {\doibase 10.1103/PhysRevB.85.060402} {\bibfield
  {journal} {\bibinfo  {journal} {{Phys. Rev. B}}\ }\textbf {\bibinfo {volume}
  {85}},\ \bibinfo {pages} {060402} (\bibinfo {year} {2012})}\BibitemShut
  {NoStop}%
\bibitem [{\citenamefont {Varney}\ \emph {et~al.}(2011)\citenamefont {Varney},
  \citenamefont {Sun}, \citenamefont {Galitski},\ and\ \citenamefont
  {Rigol}}]{Varney2011}%
  \BibitemOpen
  \bibfield  {author} {\bibinfo {author} {\bibfnamefont {C.~N.}\ \bibnamefont
  {Varney}}, \bibinfo {author} {\bibfnamefont {K.}~\bibnamefont {Sun}},
  \bibinfo {author} {\bibfnamefont {V.}~\bibnamefont {Galitski}}, \ and\
  \bibinfo {author} {\bibfnamefont {M.}~\bibnamefont {Rigol}},\ }\href
  {\doibase 10.1103/PhysRevLett.107.077201} {\bibfield  {journal} {\bibinfo
  {journal} {{Phys. Rev. Lett.}}\ }\textbf {\bibinfo {volume} {107}},\ \bibinfo
  {pages} {077201} (\bibinfo {year} {2011})}\BibitemShut {NoStop}%
\bibitem [{\citenamefont {Micheli}\ \emph {et~al.}(2006)\citenamefont
  {Micheli}, \citenamefont {Brennen},\ and\ \citenamefont
  {Zoller}}]{Micheli2006}%
  \BibitemOpen
  \bibfield  {author} {\bibinfo {author} {\bibfnamefont {A.}~\bibnamefont
  {Micheli}}, \bibinfo {author} {\bibfnamefont {G.~K.}\ \bibnamefont
  {Brennen}}, \ and\ \bibinfo {author} {\bibfnamefont {P.}~\bibnamefont
  {Zoller}},\ }\href {\doibase 10.1038/nphys287} {\bibfield  {journal}
  {\bibinfo  {journal} {{Nat Phys}}\ }\textbf {\bibinfo {volume} {2}},\
  \bibinfo {pages} {341} (\bibinfo {year} {2006})}\BibitemShut {NoStop}%
\bibitem [{\citenamefont {Gorshkov}\ \emph {et~al.}(2013)\citenamefont
  {Gorshkov}, \citenamefont {Hazzard},\ and\ \citenamefont
  {Rey}}]{Gorshkov2013}%
  \BibitemOpen
  \bibfield  {author} {\bibinfo {author} {\bibfnamefont {A.~V.}\ \bibnamefont
  {Gorshkov}}, \bibinfo {author} {\bibfnamefont {K.~R.}\ \bibnamefont
  {Hazzard}}, \ and\ \bibinfo {author} {\bibfnamefont {A.~M.}\ \bibnamefont
  {Rey}},\ }\href {\doibase 10.1080/00268976.2013.800604} {\bibfield  {journal}
  {\bibinfo  {journal} {{Molecular Physics}}\ }\textbf {\bibinfo {volume}
  {111}},\ \bibinfo {pages} {1908} (\bibinfo {year} {2013})}\BibitemShut
  {NoStop}%
\bibitem [{\citenamefont {Sherkunov}\ \emph {et~al.}(2012)\citenamefont
  {Sherkunov}, \citenamefont {Cheianov},\ and\ \citenamefont
  {Fal'ko}}]{Sherkunov2012}%
  \BibitemOpen
  \bibfield  {author} {\bibinfo {author} {\bibfnamefont {Y.}~\bibnamefont
  {Sherkunov}}, \bibinfo {author} {\bibfnamefont {V.~V.}\ \bibnamefont
  {Cheianov}}, \ and\ \bibinfo {author} {\bibfnamefont {V.}~\bibnamefont
  {Fal'ko}},\ }\href {\doibase 10.1103/PhysRevA.85.025603} {\bibfield
  {journal} {\bibinfo  {journal} {{Phys. Rev. A}}\ }\textbf {\bibinfo {volume}
  {85}},\ \bibinfo {pages} {025603} (\bibinfo {year} {2012})}\BibitemShut
  {NoStop}%
\bibitem [{\citenamefont {Knap}\ \emph {et~al.}(2012)\citenamefont {Knap},
  \citenamefont {Berg}, \citenamefont {Ganahl},\ and\ \citenamefont
  {Demler}}]{Knap2012}%
  \BibitemOpen
  \bibfield  {author} {\bibinfo {author} {\bibfnamefont {M.}~\bibnamefont
  {Knap}}, \bibinfo {author} {\bibfnamefont {E.}~\bibnamefont {Berg}}, \bibinfo
  {author} {\bibfnamefont {M.}~\bibnamefont {Ganahl}}, \ and\ \bibinfo {author}
  {\bibfnamefont {E.}~\bibnamefont {Demler}},\ }\href {\doibase
  10.1103/PhysRevB.86.064501} {\bibfield  {journal} {\bibinfo  {journal}
  {{Phys. Rev. B}}\ }\textbf {\bibinfo {volume} {86}},\ \bibinfo {pages}
  {064501} (\bibinfo {year} {2012})}\BibitemShut {NoStop}%
\bibitem [{\citenamefont {Volosniev}\ \emph {et~al.}(2013)\citenamefont
  {Volosniev}, \citenamefont {Armstrong}, \citenamefont {Fedorov},
  \citenamefont {Jensen}, \citenamefont {Valiente},\ and\ \citenamefont
  {Zinner}}]{Volosniev2013}%
  \BibitemOpen
  \bibfield  {author} {\bibinfo {author} {\bibfnamefont {A.~G.}\ \bibnamefont
  {Volosniev}}, \bibinfo {author} {\bibfnamefont {J.~R.}\ \bibnamefont
  {Armstrong}}, \bibinfo {author} {\bibfnamefont {D.~V.}\ \bibnamefont
  {Fedorov}}, \bibinfo {author} {\bibfnamefont {A.~S.}\ \bibnamefont {Jensen}},
  \bibinfo {author} {\bibfnamefont {M.}~\bibnamefont {Valiente}}, \ and\
  \bibinfo {author} {\bibfnamefont {N.~T.}\ \bibnamefont {Zinner}},\ }\href
  {\doibase 10.1088/1367-2630/15/4/043046} {\bibfield  {journal} {\bibinfo
  {journal} {{New J. Phys.}}\ }\textbf {\bibinfo {volume} {15}},\ \bibinfo
  {pages} {043046} (\bibinfo {year} {2013})}\BibitemShut {NoStop}%
\bibitem [{\citenamefont {Bauer}\ and\ \citenamefont
  {Parish}(2012)}]{Bauer2012}%
  \BibitemOpen
  \bibfield  {author} {\bibinfo {author} {\bibfnamefont {M.}~\bibnamefont
  {Bauer}}\ and\ \bibinfo {author} {\bibfnamefont {M.~M.}\ \bibnamefont
  {Parish}},\ }\href {\doibase 10.1103/PhysRevLett.108.255302} {\bibfield
  {journal} {\bibinfo  {journal} {{Phys. Rev. Lett.}}\ }\textbf {\bibinfo
  {volume} {108}},\ \bibinfo {pages} {255302} (\bibinfo {year}
  {2012})}\BibitemShut {NoStop}%
\bibitem [{\citenamefont {Trotzky}\ \emph {et~al.}(2011)\citenamefont
  {Trotzky}, \citenamefont {Chen}, \citenamefont {Flesch}, \citenamefont
  {McCulloch}, \citenamefont {Schollw\"{o}ck}, \citenamefont {Eisert},\ and\
  \citenamefont {Bloch}}]{Trotzky2011c}%
  \BibitemOpen
  \bibfield  {author} {\bibinfo {author} {\bibfnamefont {S.}~\bibnamefont
  {Trotzky}}, \bibinfo {author} {\bibfnamefont {Y.-A.}\ \bibnamefont {Chen}},
  \bibinfo {author} {\bibfnamefont {A.}~\bibnamefont {Flesch}}, \bibinfo
  {author} {\bibfnamefont {I.~P.}\ \bibnamefont {McCulloch}}, \bibinfo {author}
  {\bibfnamefont {U.}~\bibnamefont {Schollw\"{o}ck}}, \bibinfo {author}
  {\bibfnamefont {J.}~\bibnamefont {Eisert}}, \ and\ \bibinfo {author}
  {\bibfnamefont {I.}~\bibnamefont {Bloch}},\ }\href {\doibase
  10.1038/nphys2232} {\bibfield  {journal} {\bibinfo  {journal} {Nature
  Physics}\ }\textbf {\bibinfo {volume} {8}},\ \bibinfo {pages} {8} (\bibinfo
  {year} {2011})},\ \Eprint {http://arxiv.org/abs/1101.2659} {arXiv:1101.2659}
  \BibitemShut {NoStop}%
\bibitem [{\citenamefont {Hofferberth}\ \emph {et~al.}(2007)\citenamefont
  {Hofferberth}, \citenamefont {Lesanovsky}, \citenamefont {Fischer},
  \citenamefont {Schumm},\ and\ \citenamefont
  {Schmiedmayer}}]{Hofferberth2007a}%
  \BibitemOpen
  \bibfield  {author} {\bibinfo {author} {\bibfnamefont {S.}~\bibnamefont
  {Hofferberth}}, \bibinfo {author} {\bibfnamefont {I.}~\bibnamefont
  {Lesanovsky}}, \bibinfo {author} {\bibfnamefont {B.}~\bibnamefont {Fischer}},
  \bibinfo {author} {\bibfnamefont {T.}~\bibnamefont {Schumm}}, \ and\ \bibinfo
  {author} {\bibfnamefont {J.}~\bibnamefont {Schmiedmayer}},\ }\href {\doibase
  10.1038/nature06149} {\bibfield  {journal} {\bibinfo  {journal} {Nature}\
  }\textbf {\bibinfo {volume} {449}},\ \bibinfo {pages} {324} (\bibinfo {year}
  {2007})},\ \Eprint {http://arxiv.org/abs/0706.2259} {arXiv:0706.2259}
  \BibitemShut {NoStop}%
\bibitem [{\citenamefont {Japaridze}\ \emph {et~al.}(2007)\citenamefont
  {Japaridze}, \citenamefont {Noack}, \citenamefont {Baeriswyl},\ and\
  \citenamefont {Tincani}}]{Japaridze2007}%
  \BibitemOpen
  \bibfield  {author} {\bibinfo {author} {\bibfnamefont {G.~I.}\ \bibnamefont
  {Japaridze}}, \bibinfo {author} {\bibfnamefont {R.~M.}\ \bibnamefont
  {Noack}}, \bibinfo {author} {\bibfnamefont {D.}~\bibnamefont {Baeriswyl}}, \
  and\ \bibinfo {author} {\bibfnamefont {L.}~\bibnamefont {Tincani}},\ }\href
  {\doibase 10.1103/PhysRevB.76.115118} {\bibfield  {journal} {\bibinfo
  {journal} {{Phys. Rev. B}}\ }\textbf {\bibinfo {volume} {76}},\ \bibinfo
  {pages} {115118} (\bibinfo {year} {2007})}\BibitemShut {NoStop}%
\bibitem [{\citenamefont {Mishra}\ \emph {et~al.}(2013)\citenamefont {Mishra},
  \citenamefont {Pai}, \citenamefont {Mukerjee},\ and\ \citenamefont
  {Paramekanti}}]{Mishra2013}%
  \BibitemOpen
  \bibfield  {author} {\bibinfo {author} {\bibfnamefont {T.}~\bibnamefont
  {Mishra}}, \bibinfo {author} {\bibfnamefont {R.~V.}\ \bibnamefont {Pai}},
  \bibinfo {author} {\bibfnamefont {S.}~\bibnamefont {Mukerjee}}, \ and\
  \bibinfo {author} {\bibfnamefont {A.}~\bibnamefont {Paramekanti}},\ }\href
  {\doibase 10.1103/PhysRevB.87.174504} {\bibfield  {journal} {\bibinfo
  {journal} {{Phys. Rev. B}}\ }\textbf {\bibinfo {volume} {87}},\ \bibinfo
  {pages} {174504} (\bibinfo {year} {2013})}\BibitemShut {NoStop}%
\bibitem [{\citenamefont {Gammelmark}\ and\ \citenamefont
  {Zinner}(2013)}]{Gammelmark2013}%
  \BibitemOpen
  \bibfield  {author} {\bibinfo {author} {\bibfnamefont {S.}~\bibnamefont
  {Gammelmark}}\ and\ \bibinfo {author} {\bibfnamefont {N.~T.}\ \bibnamefont
  {Zinner}},\ }\href {\doibase 10.1103/PhysRevB.88.245135} {\bibfield
  {journal} {\bibinfo  {journal} {{Phys. Rev. B}}\ }\textbf {\bibinfo {volume}
  {88}},\ \bibinfo {pages} {245135} (\bibinfo {year} {2013})}\BibitemShut
  {NoStop}%
\bibitem [{\citenamefont {Ni}\ \emph {et~al.}(2008)\citenamefont {Ni},
  \citenamefont {Ospelkaus}, \citenamefont {Miranda}, \citenamefont {Pe'er},
  \citenamefont {Neyenhuis}, \citenamefont {Zirbel}, \citenamefont
  {Kotochigova}, \citenamefont {Julienne}, \citenamefont {Jin},\ and\
  \citenamefont {Ye}}]{Ni2008}%
  \BibitemOpen
  \bibfield  {author} {\bibinfo {author} {\bibfnamefont {K.-K.}\ \bibnamefont
  {Ni}}, \bibinfo {author} {\bibfnamefont {S.}~\bibnamefont {Ospelkaus}},
  \bibinfo {author} {\bibfnamefont {M.~H. G.~d.}\ \bibnamefont {Miranda}},
  \bibinfo {author} {\bibfnamefont {A.}~\bibnamefont {Pe'er}}, \bibinfo
  {author} {\bibfnamefont {B.}~\bibnamefont {Neyenhuis}}, \bibinfo {author}
  {\bibfnamefont {J.~J.}\ \bibnamefont {Zirbel}}, \bibinfo {author}
  {\bibfnamefont {S.}~\bibnamefont {Kotochigova}}, \bibinfo {author}
  {\bibfnamefont {P.~S.}\ \bibnamefont {Julienne}}, \bibinfo {author}
  {\bibfnamefont {D.~S.}\ \bibnamefont {Jin}}, \ and\ \bibinfo {author}
  {\bibfnamefont {J.}~\bibnamefont {Ye}},\ }\href {\doibase
  10.1126/science.1163861} {\bibfield  {journal} {\bibinfo  {journal}
  {{Science}}\ }\textbf {\bibinfo {volume} {322}},\ \bibinfo {pages} {231}
  (\bibinfo {year} {2008})},\ \bibinfo {note} {pMID: 18801969}\BibitemShut
  {NoStop}%
\bibitem [{\citenamefont {Takekoshi}\ \emph {et~al.}(2012)\citenamefont
  {Takekoshi}, \citenamefont {Debatin}, \citenamefont {Rameshan}, \citenamefont
  {Ferlaino}, \citenamefont {Grimm}, \citenamefont {N\"{a}gerl}, \citenamefont
  {Le~Sueur}, \citenamefont {Hutson}, \citenamefont {Julienne}, \citenamefont
  {Kotochigova},\ and\ \citenamefont {Tiemann}}]{Takekoshi2012}%
  \BibitemOpen
  \bibfield  {author} {\bibinfo {author} {\bibfnamefont {T.}~\bibnamefont
  {Takekoshi}}, \bibinfo {author} {\bibfnamefont {M.}~\bibnamefont {Debatin}},
  \bibinfo {author} {\bibfnamefont {R.}~\bibnamefont {Rameshan}}, \bibinfo
  {author} {\bibfnamefont {F.}~\bibnamefont {Ferlaino}}, \bibinfo {author}
  {\bibfnamefont {R.}~\bibnamefont {Grimm}}, \bibinfo {author} {\bibfnamefont
  {H.-C.}\ \bibnamefont {N\"{a}gerl}}, \bibinfo {author} {\bibfnamefont
  {C.~R.}\ \bibnamefont {Le~Sueur}}, \bibinfo {author} {\bibfnamefont {J.~M.}\
  \bibnamefont {Hutson}}, \bibinfo {author} {\bibfnamefont {P.~S.}\
  \bibnamefont {Julienne}}, \bibinfo {author} {\bibfnamefont {S.}~\bibnamefont
  {Kotochigova}}, \ and\ \bibinfo {author} {\bibfnamefont {E.}~\bibnamefont
  {Tiemann}},\ }\href {\doibase 10.1103/PhysRevA.85.032506} {\bibfield
  {journal} {\bibinfo  {journal} {{Phys. Rev. A}}\ }\textbf {\bibinfo {volume}
  {85}},\ \bibinfo {pages} {032506} (\bibinfo {year} {2012})}\BibitemShut
  {NoStop}%
\bibitem [{\citenamefont {Yan}\ \emph {et~al.}(2013)\citenamefont {Yan},
  \citenamefont {Moses}, \citenamefont {Gadway}, \citenamefont {Covey},
  \citenamefont {Hazzard}, \citenamefont {Rey}, \citenamefont {Jin},\ and\
  \citenamefont {Ye}}]{Yan2013}%
  \BibitemOpen
  \bibfield  {author} {\bibinfo {author} {\bibfnamefont {B.}~\bibnamefont
  {Yan}}, \bibinfo {author} {\bibfnamefont {S.~A.}\ \bibnamefont {Moses}},
  \bibinfo {author} {\bibfnamefont {B.}~\bibnamefont {Gadway}}, \bibinfo
  {author} {\bibfnamefont {J.~P.}\ \bibnamefont {Covey}}, \bibinfo {author}
  {\bibfnamefont {K.~R.~A.}\ \bibnamefont {Hazzard}}, \bibinfo {author}
  {\bibfnamefont {A.~M.}\ \bibnamefont {Rey}}, \bibinfo {author} {\bibfnamefont
  {D.~S.}\ \bibnamefont {Jin}}, \ and\ \bibinfo {author} {\bibfnamefont
  {J.}~\bibnamefont {Ye}},\ }\href {\doibase 10.1038/nature12483} {\bibfield
  {journal} {\bibinfo  {journal} {{Nature}}\ }\textbf {\bibinfo {volume}
  {501}},\ \bibinfo {pages} {521} (\bibinfo {year} {2013})}\BibitemShut
  {NoStop}%
\bibitem [{\citenamefont {Lahaye}\ \emph {et~al.}(2007)\citenamefont {Lahaye},
  \citenamefont {Koch}, \citenamefont {Fr\"{o}hlich}, \citenamefont {Fattori},
  \citenamefont {Metz}, \citenamefont {Griesmaier}, \citenamefont
  {Giovanazzi},\ and\ \citenamefont {Pfau}}]{Lahaye2007}%
  \BibitemOpen
  \bibfield  {author} {\bibinfo {author} {\bibfnamefont {T.}~\bibnamefont
  {Lahaye}}, \bibinfo {author} {\bibfnamefont {T.}~\bibnamefont {Koch}},
  \bibinfo {author} {\bibfnamefont {B.}~\bibnamefont {Fr\"{o}hlich}}, \bibinfo
  {author} {\bibfnamefont {M.}~\bibnamefont {Fattori}}, \bibinfo {author}
  {\bibfnamefont {J.}~\bibnamefont {Metz}}, \bibinfo {author} {\bibfnamefont
  {A.}~\bibnamefont {Griesmaier}}, \bibinfo {author} {\bibfnamefont
  {S.}~\bibnamefont {Giovanazzi}}, \ and\ \bibinfo {author} {\bibfnamefont
  {T.}~\bibnamefont {Pfau}},\ }\href {\doibase 10.1038/nature06036} {\bibfield
  {journal} {\bibinfo  {journal} {{Nature}}\ }\textbf {\bibinfo {volume}
  {448}},\ \bibinfo {pages} {672} (\bibinfo {year} {2007})}\BibitemShut
  {NoStop}%
\bibitem [{\citenamefont {Chotia}\ \emph {et~al.}(2012)\citenamefont {Chotia},
  \citenamefont {Neyenhuis}, \citenamefont {Moses}, \citenamefont {Yan},
  \citenamefont {Covey}, \citenamefont {Foss-Feig}, \citenamefont {Rey},
  \citenamefont {Jin},\ and\ \citenamefont {Ye}}]{Chotia2012}%
  \BibitemOpen
  \bibfield  {author} {\bibinfo {author} {\bibfnamefont {A.}~\bibnamefont
  {Chotia}}, \bibinfo {author} {\bibfnamefont {B.}~\bibnamefont {Neyenhuis}},
  \bibinfo {author} {\bibfnamefont {S.~A.}\ \bibnamefont {Moses}}, \bibinfo
  {author} {\bibfnamefont {B.}~\bibnamefont {Yan}}, \bibinfo {author}
  {\bibfnamefont {J.~P.}\ \bibnamefont {Covey}}, \bibinfo {author}
  {\bibfnamefont {M.}~\bibnamefont {Foss-Feig}}, \bibinfo {author}
  {\bibfnamefont {A.~M.}\ \bibnamefont {Rey}}, \bibinfo {author} {\bibfnamefont
  {D.~S.}\ \bibnamefont {Jin}}, \ and\ \bibinfo {author} {\bibfnamefont
  {J.}~\bibnamefont {Ye}},\ }\href {\doibase 10.1103/PhysRevLett.108.080405}
  {\bibfield  {journal} {\bibinfo  {journal} {{Phys. Rev. Lett.}}\ }\textbf
  {\bibinfo {volume} {108}},\ \bibinfo {pages} {080405} (\bibinfo {year}
  {2012})}\BibitemShut {NoStop}%
\bibitem [{\citenamefont {Lieb}\ and\ \citenamefont
  {Liniger}(1963)}]{Lieb1963}%
  \BibitemOpen
  \bibfield  {author} {\bibinfo {author} {\bibfnamefont {E.~H.}\ \bibnamefont
  {Lieb}}\ and\ \bibinfo {author} {\bibfnamefont {W.}~\bibnamefont {Liniger}},\
  }\href {\doibase 10.1103/PhysRev.130.1605} {\bibfield  {journal} {\bibinfo
  {journal} {Physical Review}\ }\textbf {\bibinfo {volume} {130}},\ \bibinfo
  {pages} {1605} (\bibinfo {year} {1963})}\BibitemShut {NoStop}%
\bibitem [{\citenamefont {Paredes}\ \emph {et~al.}(2004)\citenamefont
  {Paredes}, \citenamefont {Widera}, \citenamefont {Murg}, \citenamefont
  {Mandel}, \citenamefont {Folling}, \citenamefont {Cirac}, \citenamefont
  {Shlyapnikov}, \citenamefont {Hansch},\ and\ \citenamefont
  {Bloch}}]{Paredes2004a}%
  \BibitemOpen
  \bibfield  {author} {\bibinfo {author} {\bibfnamefont {B.}~\bibnamefont
  {Paredes}}, \bibinfo {author} {\bibfnamefont {A.}~\bibnamefont {Widera}},
  \bibinfo {author} {\bibfnamefont {V.}~\bibnamefont {Murg}}, \bibinfo {author}
  {\bibfnamefont {O.}~\bibnamefont {Mandel}}, \bibinfo {author} {\bibfnamefont
  {S.}~\bibnamefont {Folling}}, \bibinfo {author} {\bibfnamefont
  {I.}~\bibnamefont {Cirac}}, \bibinfo {author} {\bibfnamefont {G.~V.}\
  \bibnamefont {Shlyapnikov}}, \bibinfo {author} {\bibfnamefont {T.~W.}\
  \bibnamefont {Hansch}}, \ and\ \bibinfo {author} {\bibfnamefont
  {I.}~\bibnamefont {Bloch}},\ }\href {\doibase 10.1038/nature02578.}
  {\bibfield  {journal} {\bibinfo  {journal} {Nature}\ }\textbf {\bibinfo
  {volume} {429}},\ \bibinfo {pages} {277} (\bibinfo {year}
  {2004})}\BibitemShut {NoStop}%
\bibitem [{\citenamefont {Garc\'{\i}a-Ripoll}\ \emph
  {et~al.}(2009)\citenamefont {Garc\'{\i}a-Ripoll}, \citenamefont {D\"{u}rr},
  \citenamefont {Syassen}, \citenamefont {Bauer}, \citenamefont {Lettner},
  \citenamefont {Rempe},\ and\ \citenamefont {Cirac}}]{GarciaRipoll2009a}%
  \BibitemOpen
  \bibfield  {author} {\bibinfo {author} {\bibfnamefont {J.~J.}\ \bibnamefont
  {Garc\'{\i}a-Ripoll}}, \bibinfo {author} {\bibfnamefont {S.}~\bibnamefont
  {D\"{u}rr}}, \bibinfo {author} {\bibfnamefont {N.}~\bibnamefont {Syassen}},
  \bibinfo {author} {\bibfnamefont {D.~M.}\ \bibnamefont {Bauer}}, \bibinfo
  {author} {\bibfnamefont {M.}~\bibnamefont {Lettner}}, \bibinfo {author}
  {\bibfnamefont {G.}~\bibnamefont {Rempe}}, \ and\ \bibinfo {author}
  {\bibfnamefont {J.~I.}\ \bibnamefont {Cirac}},\ }\href {\doibase
  10.1088/1367-2630/11/1/013053} {\bibfield  {journal} {\bibinfo  {journal}
  {New Journal of Physics}\ }\textbf {\bibinfo {volume} {11}} (\bibinfo {year}
  {2009}),\ 10.1088/1367-2630/11/1/013053}\BibitemShut {NoStop}%
\bibitem [{\citenamefont {Mikeska}\ and\ \citenamefont
  {Kolezhuk}(2004)}]{Mikeska2004}%
  \BibitemOpen
  \bibfield  {author} {\bibinfo {author} {\bibfnamefont {H.-J.}\ \bibnamefont
  {Mikeska}}\ and\ \bibinfo {author} {\bibfnamefont {A.~K.}\ \bibnamefont
  {Kolezhuk}},\ }in\ \href
  {http://link.springer.com/chapter/10.1007/BFb0119591} {\emph {\bibinfo
  {booktitle} {{Quantum Magnetism}}}},\ \bibinfo {series and number} {\bibinfo
  {series} {Lecture Notes in Physics}\ No.\ \bibinfo {number} {645}},\ \bibinfo
  {editor} {edited by\ \bibinfo {editor} {\bibfnamefont {U.}~\bibnamefont
  {Schollw\"{o}ck}}, \bibinfo {editor} {\bibfnamefont {J.}~\bibnamefont
  {Richter}}, \bibinfo {editor} {\bibfnamefont {D.~J.~J.}\ \bibnamefont
  {Farnell}}, \ and\ \bibinfo {editor} {\bibfnamefont {R.~F.}\ \bibnamefont
  {Bishop}}}\ (\bibinfo  {publisher} {{Springer Berlin Heidelberg}},\ \bibinfo
  {year} {2004})\ pp.\ \bibinfo {pages} {1--83}\BibitemShut {NoStop}%
\bibitem [{\citenamefont {Giamarchi}(2004)}]{Giamarchi2004}%
  \BibitemOpen
  \bibfield  {author} {\bibinfo {author} {\bibfnamefont {T.}~\bibnamefont
  {Giamarchi}},\ }\href@noop {} {\emph {\bibinfo {title} {Quantum Physics in
  One Dimension}}}\ (\bibinfo  {publisher} {{Oxford University Press}},\
  \bibinfo {year} {2004})\BibitemShut {NoStop}%
\bibitem [{\citenamefont {Orignac}\ and\ \citenamefont
  {Giamarchi}(1998)}]{Orignac1998}%
  \BibitemOpen
  \bibfield  {author} {\bibinfo {author} {\bibfnamefont {E.}~\bibnamefont
  {Orignac}}\ and\ \bibinfo {author} {\bibfnamefont {T.}~\bibnamefont
  {Giamarchi}},\ }\href {\doibase 10.1103/PhysRevB.57.5812} {\bibfield
  {journal} {\bibinfo  {journal} {{Phys. Rev. B}}\ }\textbf {\bibinfo {volume}
  {57}},\ \bibinfo {pages} {5812} (\bibinfo {year} {1998})}\BibitemShut
  {NoStop}%
\bibitem [{\citenamefont {Cross}(1979)}]{Cross1979}%
  \BibitemOpen
  \bibfield  {author} {\bibinfo {author} {\bibfnamefont {M.}~\bibnamefont
  {Cross}},\ }\href {\doibase 10.1103/PhysRevB.19.402} {\bibfield  {journal}
  {\bibinfo  {journal} {{Phys. Rev. B}}\ }\textbf {\bibinfo {volume} {19}},\
  \bibinfo {pages} {402} (\bibinfo {year} {1979})}\BibitemShut {NoStop}%
\bibitem [{\citenamefont {Orignac}(2004)}]{Orignac2004}%
  \BibitemOpen
  \bibfield  {author} {\bibinfo {author} {\bibfnamefont {E.}~\bibnamefont
  {Orignac}},\ }\href {\doibase 10.1103/PhysRevB.70.214436} {\bibfield
  {journal} {\bibinfo  {journal} {{Phys. Rev. B}}\ }\textbf {\bibinfo {volume}
  {70}} (\bibinfo {year} {2004}),\ 10.1103/PhysRevB.70.214436}\BibitemShut
  {NoStop}%
\bibitem [{\citenamefont {Sirker}(2012)}]{Sirker2012}%
  \BibitemOpen
  \bibfield  {author} {\bibinfo {author} {\bibfnamefont {J.}~\bibnamefont
  {Sirker}},\ }\href {\doibase 10.1142/S0217979212440092} {\bibfield  {journal}
  {\bibinfo  {journal} {{International Journal of Modern Physics B}}\ }\textbf
  {\bibinfo {volume} {26}},\ \bibinfo {pages} {1244009} (\bibinfo {year}
  {2012})}\BibitemShut {NoStop}%
\bibitem [{\citenamefont {Furukawa}\ \emph {et~al.}(2012)\citenamefont
  {Furukawa}, \citenamefont {Sato}, \citenamefont {Onoda},\ and\ \citenamefont
  {Furusaki}}]{Furukawa2012a}%
  \BibitemOpen
  \bibfield  {author} {\bibinfo {author} {\bibfnamefont {S.}~\bibnamefont
  {Furukawa}}, \bibinfo {author} {\bibfnamefont {M.}~\bibnamefont {Sato}},
  \bibinfo {author} {\bibfnamefont {S.}~\bibnamefont {Onoda}}, \ and\ \bibinfo
  {author} {\bibfnamefont {A.}~\bibnamefont {Furusaki}},\ }\href {\doibase
  10.1103/PhysRevB.86.094417} {\bibfield  {journal} {\bibinfo  {journal}
  {Physical Review B - Condensed Matter and Materials Physics}\ }\textbf
  {\bibinfo {volume} {86}} (\bibinfo {year} {2012}),\
  10.1103/PhysRevB.86.094417},\ \Eprint
  {http://arxiv.org/abs/arXiv:1207.1059v1} {arXiv:arXiv:1207.1059v1}
  \BibitemShut {NoStop}%
\bibitem [{\citenamefont {Furukawa}\ \emph {et~al.}(2010)\citenamefont
  {Furukawa}, \citenamefont {Sato},\ and\ \citenamefont
  {Furusaki}}]{Furukawa2010b}%
  \BibitemOpen
  \bibfield  {author} {\bibinfo {author} {\bibfnamefont {S.}~\bibnamefont
  {Furukawa}}, \bibinfo {author} {\bibfnamefont {M.}~\bibnamefont {Sato}}, \
  and\ \bibinfo {author} {\bibfnamefont {A.}~\bibnamefont {Furusaki}},\ }\href
  {\doibase 10.1103/PhysRevB.81.094430} {\bibfield  {journal} {\bibinfo
  {journal} {Physical Review B - Condensed Matter and Materials Physics}\
  }\textbf {\bibinfo {volume} {81}},\ \bibinfo {pages} {1} (\bibinfo {year}
  {2010})},\ \Eprint {http://arxiv.org/abs/1003.4517} {arXiv:1003.4517}
  \BibitemShut {NoStop}%
\bibitem [{\citenamefont {Burnell}\ \emph {et~al.}(2009)\citenamefont
  {Burnell}, \citenamefont {Parish}, \citenamefont {Cooper},\ and\
  \citenamefont {Sondhi}}]{Burnell2009}%
  \BibitemOpen
  \bibfield  {author} {\bibinfo {author} {\bibfnamefont {F.~J.}\ \bibnamefont
  {Burnell}}, \bibinfo {author} {\bibfnamefont {M.~M.}\ \bibnamefont {Parish}},
  \bibinfo {author} {\bibfnamefont {N.~R.}\ \bibnamefont {Cooper}}, \ and\
  \bibinfo {author} {\bibfnamefont {S.~L.}\ \bibnamefont {Sondhi}},\ }\href
  {\doibase 10.1103/PhysRevB.80.174519} {\bibfield  {journal} {\bibinfo
  {journal} {{Phys. Rev. B}}\ }\textbf {\bibinfo {volume} {80}},\ \bibinfo
  {pages} {174519} (\bibinfo {year} {2009})}\BibitemShut {NoStop}%
\end{thebibliography}%

\end{document}